\documentclass[a4paper,12pt,useAMS]{mn2e}

\usepackage{graphicx}
\usepackage{amssymb}

\title[The Universal Einstein Radius Distribution]{The Universal Einstein Radius Distribution from 10,000 SDSS Clusters}
\author[Zitrin et al.]{Adi Zitrin$^{1,4}$\thanks{E-mail:adiz@wise.tau.ac.il}, Tom Broadhurst$^{2,3}$, Matthias Bartelmann$^{4}$, Yoel Rephaeli$^{1}$, \and Masamune Oguri$^{5,6}$, Narciso Ben\'itez$^{7}$, Jiangang Hao$^{8}$, Keiichi Umetsu$^{9}$\\\\\\
$^{1}$The School of Physics and Astronomy, the Raymond and Beverly Sackler Faculty of Exact Sciences, Tel Aviv University,\\ Tel Aviv 69978, Israel\\
$^{2}$Department of Theoretical Physics, University of Basque Country UPV/EHU, Leioa, Spain\\
$^{3}$IKERBASQUE, Basque Foundation for Science\\
$^{4}$Institut f\"{u}r Theoretische Astrophysik, ZAH, Albert-Ueberle-Stra\ss e 2, 69120 Heidelberg, Germany\\
$^{5}$Institute for the Physics and Mathematics of the Universe, University of Tokyo, 5-1-5 Kashiwanoha, Kashiwa, Chiba 277-8583, Japan\\
$^{6}$Division of Theoretical Astronomy, National Astronomical Observatory of Japan, 2-21-1 Osawa, Mitaka, Tokyo 181-8588, Japan\\
$^{7}$Instituto de Astrof\'isica de Andaluc\'ia (CSIC), C/Camino Bajo de Hu\'etor, 24, Granada, 18008, Spain\\
$^{8}$Center for Particle Astrophysics, Fermi National Accelerator Laboratory, Batavia, IL 60510\\
$^{9}$Institute of Astronomy and Astrophysics, Academia Sinica, P.~O. Box 23-141, Taipei 10617, Taiwan}

\begin{document}


\pagerange{\pageref{firstpage}--\pageref{lastpage}} \pubyear{2010}

\maketitle

\label{firstpage}

\begin{abstract}

We present results from strong-lens modelling of 10,000 SDSS clusters, to establish the universal distribution of Einstein radii. Detailed lensing analyses have shown that the inner mass distribution of clusters can be accurately modelled by assuming light traces mass, successfully uncovering large numbers of multiple-images. Approximate critical curves and the effective Einstein radius of each cluster can therefore be readily calculated, from the distribution of member galaxies and scaled by their luminosities. We use a subsample of 10 well-studied clusters covered by both SDSS and HST to calibrate and test this method, and show that an accurate determination of the Einstein radius and mass can be achieved by this approach ``blindly'', in an automated way, and without requiring multiple images as input. We present the results of the first 10,000 clusters analysed in the range $0.1<z<0.55$, and compare them to theoretical expectations. We find that for this all-sky representative sample the Einstein radius distribution is log-normal in shape, with $\langle Log(\theta_{e}\arcsec)\rangle=0.73^{+0.02}_{-0.03}$, $\sigma=0.316^{+0.004}_{-0.002}$, and with higher abundance of large $\theta_{e}$ clusters than predicted by $\Lambda$CDM. We visually inspect each of the clusters with $\theta_{e}>40 \arcsec$ ($z_{s}=2$) and find that $\sim20\%$ are boosted by various projection effects detailed here, remaining with $\sim40$ real giant-lens candidates, with a maximum of $\theta_{e}=69\pm12 \arcsec$ ($z_{s}=2$) for the most massive candidate, in agreement with semi-analytic calculations. The results of this work should be verified further when an extended calibration sample is available.

\end{abstract}


\section{Introduction}\label{intro}

Clusters of galaxies play a fundamental role in testing cosmological models, by virtue of their position at the high end of the cosmic mass
spectrum. Massive galaxy clusters gravitationally-lens background objects, forming distorted, magnified, and often multiple images of the same source, when the cluster surface density is high enough. These effects are in turn used to map the gravitational potentials and mass of the lensing clusters, hence providing some of the best constraints on the nature and shape of the underlying matter distributions (Broadhurst et al. 2005a, Brada\v{c} et al. 2006, Coe et al. 2010, Zitrin et al. 2010, Merten et al. 2011).

Large sky surveys such as the \emph{Sloan Digital Sky Survey} (SDSS; see Abazajian et al. 2003,2009) allow for important scientific work with different astrophysical implications (e.g., Tegmark et al. 2004, 2006, Tremonti et al. 2004, Eisenstein et al. 2005, Seljak et al. 2005, Wojtak, Hansen, \& Hjorth 2011). The large amount of data enables extensive studies with a clear statistical advantage. Here we make use of the results of a new cluster-finding algorithm operated on the SDSS DR7 data (Hao et al. 2010; on DR7 data see Abazajian et al. 2009), in order to derive the Einstein radius distribution of a significant, statistical sample. As presented in their work, more than 55,000 clusters were found using this successful and rather conservative algorithm, which we have taken upon to analyse using our improved lensing-analysis tools (e.g., Zitrin et al. 2009b, see more details in \S 2), presenting here the results of the first 10,000 clusters analysed.

The effective Einstein radius plays an important role in various studies. The Einstein radius describes the area in which multiply-lensed images may be seen due to the high mass-density of the cluster. By definition, within this critical area the average mass density is equal to $\Sigma_{crit}$ (for symmetric lenses), the critical density required for strong-lensing, whose value is dependent on the source and lens distances. In general, obtaining the critical curves with great accuracy allows matching up multiple-images, which in turn help to improve and better-constrain the model in order to derive the mass distribution and profile more accurately, teaching us about certain properties of both the observed and unseen matter. The Einstein radius therefore constitutes a measure of the strong-lens size (and efficiency), and directly enables us to estimate the amount of mass enclosed within it; $\theta_{e}=(\frac{4GM(<\theta_{e})}{c^{2}}\frac{d_{ls}}{d_{l}d_{s}})^{1/2}$ for symmetric lenses (e.g., Narayan \& Bartelmann 1996, Bartelmann 2010), where $d_{l}$, $d_{s}$, and $d_{ls}$, are the lens, source and lens-to-source (angular-diameter) distances, respectively. Equivalently, the effective Einstein radius used here is simply a measure of the critical area, $A$, so that $\theta_{e}=\sqrt{A/\pi}$.

In recent years it has been proposed that the Einstein radius distributions of several small samples of clusters, pose a challenge to $\Lambda$CDM (e.g., Broadhurst \& Barkana 2008, Zitrin et al. 2009a, 2011a). Other discrepancies such as the arc abundance, several uniquely large Einstein radii, massive high-$z$ clusters, high NFW concentration parameters, and comparison to N-body simulations, contribute further to this tension, though most studies show mainly a moderate discrepancy (e.g., Bartelmann et al. 1995, Wambsganss et al. 1995, Dalal, Holder, \& Hennawi 2004, Broadhurst et al. 2005b, 2008, Hennawi et al. 2007a,b, Hilbert et al. 2007, Sadeh \& Rephaeli 2008, Oguri \& Blandford 2009, Oguri et al. 2009, Puchwein \& Hilbert 2009, Meneghetti et al. 2010a,2011, Sereno, Jetzer \& Lubini 2010, Gralla et al. 2011, Horesh et al. 2011, Umetsu et al. 2011a, Zitrin et al. 2011a,c). Obtaining a credible empirical distribution of Einstein radii from an unprecedentedly large sample is of clear value, welcoming in addition complementary mass measurements through similarly automatic weak-lensing analyses (e.g., Hildebrandt et al. 2011) and other observations, such as of X-ray emission or the SZ-effect, when possible.

The advances in computational power over the past decades along with higher quality data and our efficient method for analysing strong-lenses (Broadhurst et al. 2005a, Zitrin et al. 2009b) now enable such an extensive study. Based on previous analyses of many clusters, we now securely determine typical physical parameters to which the critical curves are relatively indifferent, so that we extrapolate and test these assumptions to perform our analysis on the sample presented here. In particular, in this work we describe a simple and efficient method to model cluster-lenses based on the light distribution of bright cluster members, which as we have targeted to show, allows to derive the Einstein radius with sufficient accuracy, in an automated mode.

Automated surveys for lensing have been presented before, though mostly based on the observed arc properties, or relate to either galaxy-lensing scale or the weak-lensing regime (e.g., Webster, Hewett \& Irwin 1988, Cabanac et al. 2006, Mandelbaum et al. 2006, Johnston et al. 2007b, Corless \& King 2009, Marshall et al. 2009, Sheldon et al. 2009, Bayliss et al. 2011a,b, Hildebrandt et al. 2011), and have yet to produce statistically-significant results for the Einstein radius distribution directly from SL modelling. Other available SL methods, though can be successful, either require the location of many multiple-images as input or currently have too many free parameters, rendering such a ``blind'' study impossible.

The SL modelling method we implement here is based on the reasonable assumption that light approximately traces mass, which we have shown is most efficient for finding new multiple-images as the mass model is initially well constrained with sufficient resolution to derive well-approximate critical curves (see Broadhurst et al. 2005a, Zitrin et al. 2009b, 2011a,b,c, Merten et al. 2011). Recently we have tested the assumptions of this approach in Abell 1703 (Zitrin et al. 2010), by applying the non-parametric technique of Liesenborgs et al. (2006, 2007, 2009) for comparison, yielding similar results with only minor differences in the overall mass distribution and critical curves, especially where galaxies are seen since they are not included in the non-parametric technique. Independently, it has been found that SL methods based on parametric modelling, i.e., based on physical assumptions or parametrisations (for other parametric methods see, e.g., Keeton 2001, Kneib et al. 1996, Gavazzi et al. 2003, Brada\v{c} et al. 2005, Jullo et al. 2007, Halkola et al. 2008), are accurate at the level of a few percent in determining the projected inner mass (Meneghetti et al. 2010b). Clearly, non-parametric techniques and methods that are based directly on arc morphologies are also important: non-parametric techniques (e.g., Diego et al. 2005, Coe et al. 2008, Merten et al. 2009) are novel in the sense that they are assumption-free and highly flexible (e.g., Coe et al. 2010, Ponente \& Diego 2011), and methods based directly on arc morphologies yield high resolution results (see also Grillo et al. 2009). The parametric method presented here, is simply aimed to produce the critical curves in an automated way based on simple physical considerations (and thus is capable of finding multiple images as we have shown constantly before), and constitutes another important step towards the ability to deduce the lensing properties of clusters in large sky surveys in an automated way, so that we aim now to present the first observationally-deduced, universal distribution of Einstein radii.

\begin{figure}
 \begin{center}
   \includegraphics[width=90mm]{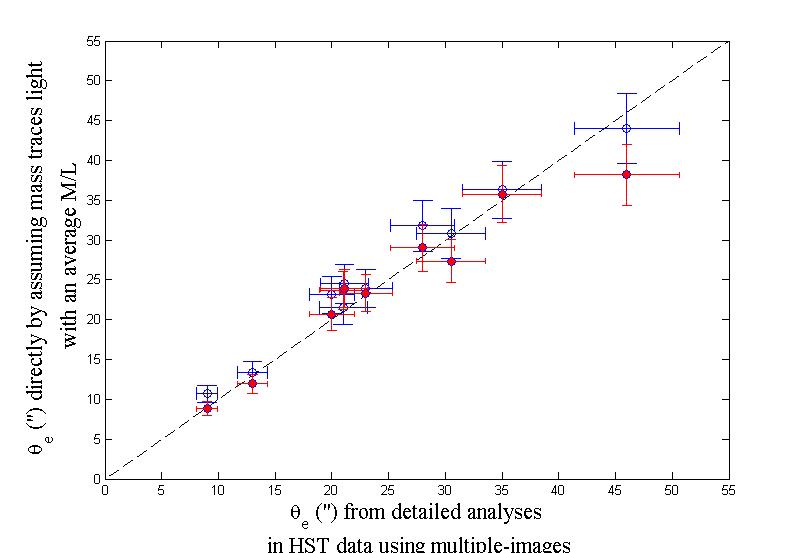}
 \end{center}
\caption{Calibration sample. Einstein radii (for $z_s=2$) derived by our ``blind'' automated algorithm in SDSS data and based on the assumption that light traces mass, with a typical $M/L$, versus the Einstein radii derived by detailed analyses of HST images of the same clusters and using the multiple images as constraints (e.g., Broadhurst et al. 2005a, Richard et al. 2010, Zitrin et al. 2010, 2011a,b). As can be seen, the ``blind'' method, based on the light distribution of bright cluster galaxies and without using any information regarding the location of multiple-images, shows remarkably similar results to those derived by the detailed independent analyses (see also Fig. \ref{comparison}). The errors in the Einstein radii are typically $\sim10\%$, and overplotted is also an $x=y$ dashed line. The comparison sample spans the redshift range $0.15<z_{l}<0.55$. \emph{Blue open circles} are the results from the blind analysis with the best-fitting parameters derived from minimising by all 10 clusters together, while \emph{red filled circles} are the results from a ``Jackknife'' minimisation described in \S \ref{jack}: in order to demonstrate how well one could assess the Einstein radius, we perform the minimisation for 9 different clusters at a time and analyse the tenth cluster with the resulting parameters. With this we obtain deviations of up to $\sim17\%$ in our ability to blindly estimate the critical curves by the automated procedure described in this work. In a complementary error-propagation check (\S \ref{jack}) we obtain similar results of $1\sigma\sim18\%$, which we take hereafter as the errors for the full-sample analysis.}
\label{TvTspec}
\end{figure}

The incorporated method involves only four free parameters. Three of them are known sufficiently well a-priori and have only negligible effect on the critical curves and resulting Einstein radius, for which we adopt typical values deduced from detailed analyses of a few dozen clusters (more details are given in \S \ref{model}). The fourth parameter, which varies from cluster to cluster, is the overall (mass) normalisation, but since the respective distances are known, this can be simply overcome by finding a typical mass-to-light ratio ($M/L$) normalisation. The $M/L$ term is embedded, in practice, in a redshift-dependent normalisation factor, which is iterated for the best fit using 10 clusters which have been accurately-analysed in HST images and have parallel SDSS data listed in the Hao et al. (2010) catalog. These include some well-known lensing clusters such as A1689, A1703, MS1358, Z2701, and others (see, e.g., Broadhurst et al. 2005a, Richard et al. 2010, Zitrin et al. 2010, 2011a,b). The results of this comparison are shown in Figure \ref{TvTspec} and Table \ref{systemo}.

The paper is organised as follows: In \S 2 we detail the modelling and the assumptions on which our algorithm is based. In \S 3 we discuss the results and relevant uncertainties, which are then summarised in \S 4. Throughout this paper we adopt a concordance $\Lambda$CDM cosmology with ($\Omega_{\rm m0}=0.3$, $\Omega_{\Lambda 0}=0.7$, $h=0.7$). All Einstein radii referred to in this work are for a fiducial source redshift of $z_{s}=2$. We also note that all logarithmic quantities in this work are in base 10, unless stated otherwise, and are denoted conventionally as ``\emph{Log}''.

\section{Strong-Lens Modelling and Analysis}\label{model}

The method we apply here is based on the simple assumption that mass traces light. This well-tested approach to lens modelling has
previously uncovered large numbers of multiply-lensed galaxies in ACS
images of e.g., Abell 1689, Cl0024, 12 high-$z$ MACS clusters, MS1358, ``Pandora's cluster'' Abell 2744, and Abell 383
(respectively, Broadhurst et al. 2005a, Zitrin et al. 2009b, 2011a,b, Merten et al. 2011, Zitrin et al. 2011c). As the basic
assumption adopted is that light approximately traces mass, the photometry of the red cluster member galaxies is used as the
starting point for the mass model.

\begin{figure}
 \begin{center}
   \includegraphics[width=85mm]{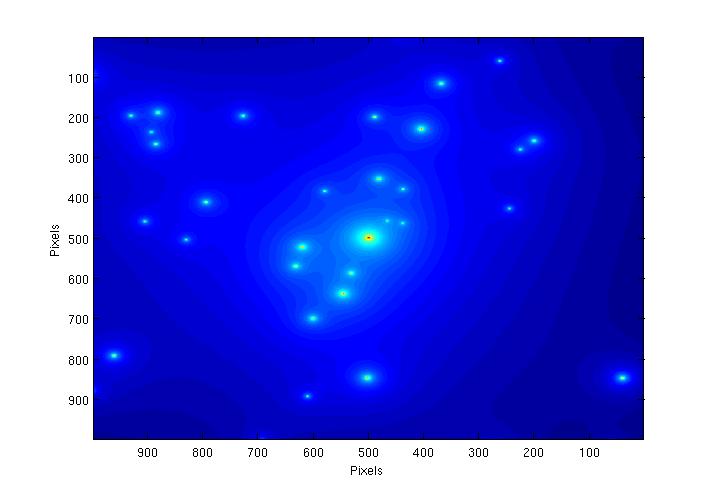}
 \end{center}
\caption{The general starting point of our lens model, where we define the
surface mass distribution based on the cluster member galaxies (see \S \ref{model}) listed in the Hao et al. (2010) cluster catalog. In this figure we show the lumpy (galaxy) component for Abell 1703 as an arbitrary example (see also Zitrin et al. 2010 for an equivalent figure but from HST observations. Axes are in pixels with $0.2 \arcsec /pixel$). We perform the same simple procedure for each of the 10,000 clusters drawn from the Hao et al. catalog.}
\label{lumpycomp}
\end{figure}

\begin{figure}
 \begin{center}
   \includegraphics[width=85mm]{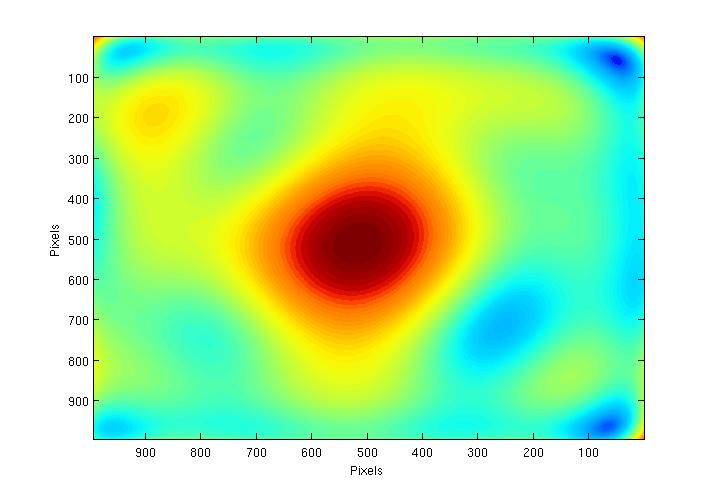}
 \end{center}
\caption{Smoothed mass distribution. To represent the DM distribution we smooth the lumpy component (Fig. \ref{lumpycomp}) of each of the 10,000 clusters drawn from Hao et al. clusters catalog (see also Zitrin et al. 2010 for an equivalent figure based on HST observations. Axes are in pixels with $0.2 \arcsec /pixel$). This smoothing procedure is most useful in generating, when combined with the lumpy component, a very reliable deflection field and corresponding critical curves, as we have shown for many clusters (e.g., Broadhurst et al. 2005a, Zitrin et al. 2009b, 2011a,b,c, Merten et al. 2011), allowing to find large numbers of multiple-images by the model.}
\label{smoothcomp}
\end{figure}

\subsection{Initial Mass Distribution}

We now wish to calculate the deflection field by the cluster galaxies, or the initial mass distribution. By assuming that the flux is proportional to the mass, i.e., assigning a certain $M/L$ ratio, the deflection field contributed by each
galaxy can now be calculated by assigning a surface-density
profile for each galaxy, $\Sigma(r)=Kr^{-q}$, which is integrated to
give the interior mass, $M(<\theta)=\frac{2\pi
K}{2-q}(d_{l}\theta)^{2-q}$. This results in a deflection angle of
(due to a single galaxy):
\begin{equation}
\label{deflection}
 \alpha(\theta)= \frac{4GM(<\theta)}{c^2\theta}\frac{d_{ls}}{d_{s}d_{l}},
\end{equation}

or more explicitly by inserting $M(<\theta)$ from above:
\begin{equation}
\label{deflectiona}
 \alpha(\theta)=\frac{4G\frac{2\pi
K}{2-q}d_{l}^{~1-q}}{c^2}\frac{d_{ls}}{d_{s}}\theta^{1-q} .
\end{equation}

We note that all quantities are known, except for $K$, the normalisation factor which is related to the M/L ratio (note that $q$ is maintained constant on a typical and known value, see \S \ref{ss:totaldef}). Thus, finding the explicit term for $K$ which scales correctly all clusters we analysed to date (taking into account the different lens and source redshifts) allows us - in principle - to perform the automated survey of Einstein radii, following the procedure described below.

By defining $K_{q}=\frac{4G}{c^{2}}\frac{2\pi
K}{2-q}\frac{d_{ls}}{d_{s}}d_{l}^{~1-q}$ we can reduce the latter formula to
get:
\begin{equation} \label{deflection2}
 \alpha(\theta)= K_{q}\theta^{1-q} ,
\end{equation}
where $K_{q}$ also depends on the redshifts involved, and on the power-law index, $q$ (which is set to constant throughout, \S \ref{ss:totaldef}).

%

The deflection angle at a certain point $\vec{\theta}$
due to the lumpy galaxy components is
simply a linear superposition of all galaxy contributions scaled by
their luminosities, $L_i$ (in $L_{\odot}$ units):
\begin{equation}
\label{deflection3}
 \vec{\alpha}_{gal}(\vec{\theta})=K_{q}\sum_{i}
\frac{L_{i}}{L_{\odot}}\, |\vec{\theta}-\vec{\theta}_i|^{1-q}
\frac{\vec{\theta}-\vec{\theta}_i}{|\vec{\theta}-\vec{\theta}_i|}.
\end{equation}

In practice we use a discretised version of equation \ref{deflection3}, over a 2D square grid $\vec\theta_m$ of $N\times N$ pixels, given by:
\begin{equation}
\label{deflection_x1}
\alpha_{gal,x}(\vec\theta_m)=K_{q}\sum_{i}
\frac{L_{i}}{L_{\odot}}\, \frac
{\Delta x_{mi}}{[(\Delta x_{mi})^2
 +
 (\Delta y_{mi})^2]^{q/2}},
 \end{equation}
\begin{equation}
\label{deflection_y1}
\alpha_{gal,y}(\vec\theta_m)=K_{q}\sum_{i}
\frac{L_{i}}{L_{\odot}}\, \frac
{\Delta y_{mi}}{[(\Delta x_{mi})^2
 +
 (\Delta y_{mi})^2]^{q/2}},
 \end{equation}
where $(\Delta x_{mi},\Delta y_{mi})$ is the displacement vector
$\vec\theta_m-\vec\theta_i$ of the $m$th pixel point, with respect to the
$i$th galaxy position $\vec\theta_i$.

Note that to obtain the luminosity $L_i$ of each member, we convert its SDSS $r$-band luminosity to the corresponding (Vega) B-band luminosity by the LRG template given in Ben\'itez et al. (2009).

From these expressions a deflection field for the galaxy contribution is easily
calculated analytically as above, and the mass distribution is now
rapidly calculated locally from the divergence of the deflection field,
i.e., the 2D equivalent of Poisson's equation. An example is given in Figure \ref{lumpycomp}.


\subsection{The Dark Matter Distribution}\label{ss:DM}

The mass contribution of galaxies is anticipated to
comprise only a small fraction of the total mass of the cluster, which
is expected to be dominated by a smooth distribution of DM. We now
simply assume that the galaxies approximately trace the DM. As mentioned, this assumption was found to work very well in earlier work on many clusters where large numbers of multiple-images were found accordingly. These multiply-lensed systems are not simply eye-ball candidates, but are reproduced and predicted by the preliminary model, indicating that this model, based on the assumption that light traces mass, is initially well constrained.

 Since the DM is of course
expected to be smoother than the distribution of galaxies, we smooth
the initial guess of the galaxy distribution obtained above, choosing for
convenience a low-order cubic spline interpolation, typical to the many previous analyses mentioned above. The smoothing degree (the polynomial degree, $S$) is also a free parameter of the model, and the deflection field contributed by the DM is then simply the sum of the contribution from each point (or pixel) in this smooth DM component. This smoothing procedure is the key to our method's success in locating multiple-images, and is in practice more useful than assuming a general DM shape such as NFW or pseudo-isothermal spheres, which are highly symmetric and do not necessarily describe the complex inner DM distribution in detail, often not allowing to find in advance the multiple images according to the initial mass distribution. An example of a smoothed component is shown in Figure \ref{smoothcomp}.

The deflection field of the DM is then (where each pixel is treated as a point mass) given by:

\begin{equation}
\label{deflection_xDM}
\alpha_{DM,x}(\vec\theta_m)=K_{q}\sum_{i}
P_i\, \frac
{\Delta x_{mi}}{[(\Delta x_{mi})^2
 +
 (\Delta y_{mi})^2]},
 \end{equation}
\begin{equation}
\label{deflection_yDM}
\alpha_{DM,y}(\vec\theta_m)=K_{q}\sum_{i}
P_i\, \frac
{\Delta y_{mi}}{[(\Delta x_{mi})^2
 +
 (\Delta y_{mi})^2]},
 \end{equation}
where $P_{i}$ represents the (unnormalised) mass value in the $i$th pixel of the smooth component. We therefore obtain now the deflection field due to the DM, hereafter
$\vec\alpha_{DM}(\vec\theta)$.

\begin{figure}[h]
 \begin{center}
   \includegraphics[width=85mm,trim=-10mm 0mm 0mm 0mm,clip]{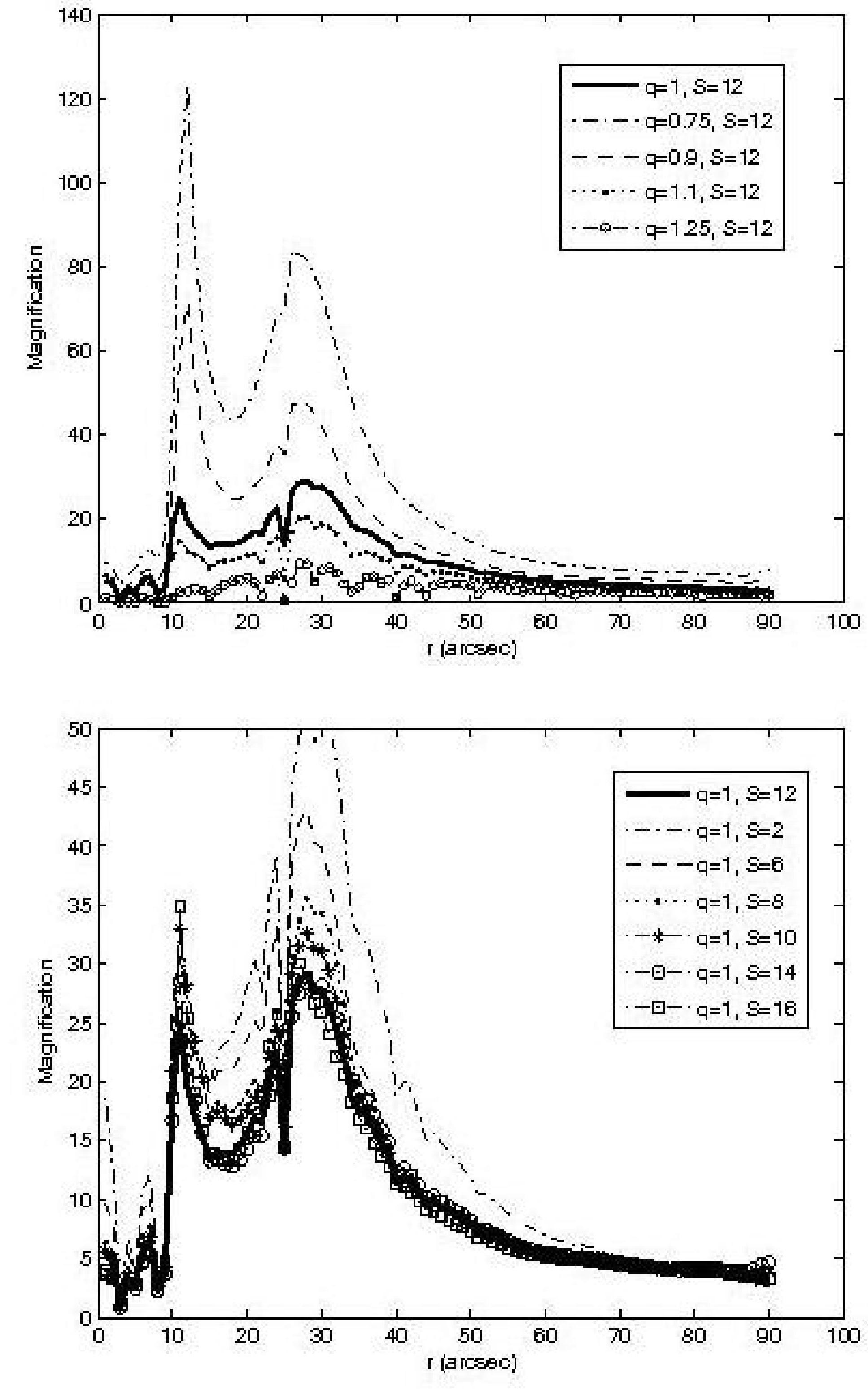}
 \end{center}
\caption{The effect of the $q$ and $S$ parameters on the location of the critical curves (or effective Einstein radius). As seen in this figure comparing different (absolute-value) magnification profiles, significant changes in both the galaxy power law, $q$ (\emph{top}), and the smoothing polynomial degree, $S$ (\emph{bottom}), affect substantially the level of magnification since these parameters control the overall profile slope, but note, these only negligibly change the locations of the critical curves or their corresponding Einstein radius, since the mass enclosed by the different mass profiles is similar within this radius. In this example, published originally in our work on Cl0024 (Zitrin et al. 2009b), the radial and tangential critical curves are at $\sim10\arcsec$ and $\sim30\arcsec$, respectively. Since here we do not wish to constrain the (magnification) profile, but only determine the location of the peaks (i.e., the critical curves), it is clear that the choice of $q$ and $S$ is not fundamentally important, since for any such combination the curves can be formed at the right location, so that for our purposes these parameters can be maintained constant on typical values. For generating this figure the critical curves were originally constrained using multiple-images (Zitrin et al. 2009b), while in this work we show that the location of the critical curves can also be reasonably constrained by using only a proper $M/L$ relation.}
\label{QSmag}
\end{figure}

\begin{figure}
 \begin{center}
   \includegraphics[width=90mm]{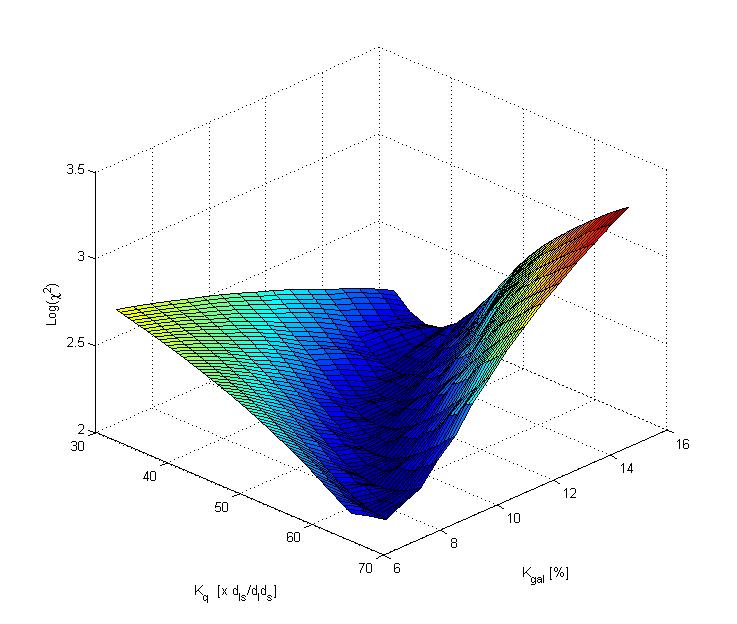}
   \includegraphics[width=90mm]{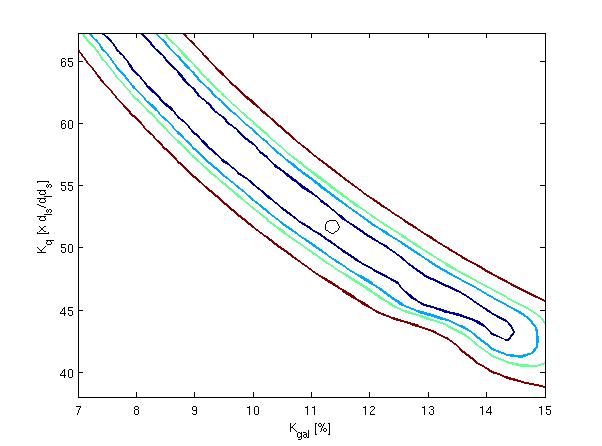}
 \end{center}
\caption{Joint $\chi^{2}$ minimisation of the relative weight of the galaxies, $K_{gal}$, and the $M/L$-dependent normalisation factor $K_{q}$, obtained by comparing the Einstein radii of the calibration sample as derived from detailed HST-based analyses, with the results of the automated analysis presented here. The \emph{top} panel shows a $\chi^{2}$ map, whereas the \emph{bottom} panel shows the 68.3\%, 95.4\%, 99\% and 99.99\% confidence levels (\emph{color countours}), and the point of best fit (\emph{black circle}). When these are jointly fit to the data, the two parameters minimised in the figure are strongly correlated, so that by fixing the relative galaxy weight, $K_{gal}$, to the best-fit constant value, we are able to reduce the number of of free parameters in our modelling to one, namely, the $M/L$-dependent normalisation factor $K_{q}$. Explicitly, this is done by fitting a least-squares line to the points within $\Delta\chi^{2}=2.3$ from the minimum $\chi^{2}$, thus obtaining the relation between the two parameters. The residuals around the minimal $\chi^{2}$ values are then used to obtain the $1\sigma$ errors. From this we obtain best fitting values (and $1\sigma$ errors) of $K_{gal}=11.4\pm0.6\%$ and $K_{q}=(51.6\pm1.9)\frac{d_{ls}}{d_{l}d_{s}}$. For more details see \S \ref{model}.}
\label{chi2figure}
\end{figure}

\subsection{The Total Deflection Field}\label{ss:totaldef}

Having calculated the two components of the deflection field, we now
simply combine them to get a total deflection field as follows:
\begin{equation}
\label{defTot}
\vec{\alpha}_T(\vec\theta)=K_{gal} \vec\alpha_{gal}(\vec\theta)+(1-K_{gal})\vec\alpha_{DM}(\vec\theta),
\end{equation}

where $K_{gal}$ is the relative contribution of the galaxy component
to the deflection field.

Both components of the deflection field are normalised by $K_{q}$, so that knowing its value enables us to approximate very well the overall deflection field. It should be stressed that although the degree of smoothing ($S$) and the index of the power-law ($q$) are the most important free parameters determining the mass profile, their effect on the Einstein radius is negligible. Based on the detailed analysis of $\sim30$ clusters (mentioned above and several more still unpublished), we note that the best-fitting parameters $q$ and $S$ show relatively little scatter among the different lenses. We can securely determine that the power-law $q$ will be in the range $1\leqslant q\leqslant1.5$, and the smoothing polynomial degree $S$ will be in the range $4\leqslant S\leqslant24$, with a sufficient resolution of $\Delta q=0.05$ and $\Delta S=2$, in order to expand the full plausible profile range per cluster. More importantly, the exact choice of $q$ and $S$ does not affect the deduced Einstein radius size, which is determined by the inner mass enclosed within it and not by the mass profile which varies (see Figs. 1 and 2 in Zitrin et al. 2009b). The crucial point to make here is that Einstein radii just constrain the enclosed mass, no matter how the mass is distributed. This is seen very clearly in Figure \ref{QSmag} here, where we show that for many different combinations of the $q$ and $S$ parameters, the critical curves form at the same radius, given a reliable constraint. For example, in Figure \ref{QSmag} the critical curves were constrained using multiple-images, while our point here, in this work, is to show that the $M/L$ ratio deduced from a calibration sample can be used as an alternative constraint, enabling an automated SL analysis. In addition, it is therefore clear, that the choice of $q$ and $S$ parameter values is not fundamentally important here, and any combination, after it is calibrated for, should in principle yield the critical curves at the right location.

With this in mind, throughout the analysis here we maintain $q$ and $S$ constants with $q$=1.2 and $S$=10, which are typical values according to our many previous analyses. With $q$ and $S$ kept constants at these values, we now constrain the fixed value of the weight of the galaxies relative to the dark matter, $K_{gal}$, and the overall normalisation factor, $K_{q}$. Having 10 clusters as a reference sample, and 2 parameters to constrain, we can well determine their values by a joint minimisation, and in turn examine how these best-fit values reproduce the reference sample critical curves. Explicitly, we perform a $\chi^2$ minimisation by comparing the Einstein radii of the calibration sample, deduced from detailed analyses based on HST observations and identification of multiple-images (see also \S \ref{intro}), with the results of the automated procedure presented here, operated on the same clusters in SDSS data:

\begin{equation}
\label{chi2}
\chi^{2}=\sum_{i}^{N}[(\theta_{e,i}^{HST}-\theta_{e,i}^{SDSS})^{2}/(\sigma_{i}^{2})],
\end{equation}

where i goes from 1 to $N=10$, for the ten calibration-sample clusters, and $\sigma_{i}$ is taken as $10\%$ of the HST deduced values, which is a typical value for the uncertainties in SL modelling results.

The results of this $\chi^2$ minimisation are seen in Figure \ref{chi2figure}. As can be seen, there is a strong correlation between the two parameters, $K_{gal}$ and $K_{q}$, which are degenerate so that many combinations of these can yield a good solution. This is a crucial point to make, since this correlation shows that indeed the number of free parameters in our modelling can be effectively reduced to one. By fixing the relative galaxy weight ($K_{gal}$) to its best-fit value, the model can now be constrained with one single parameter, namely, the $M/L$-related parameter $K_{q}$. To do this, we fit by a least-squares minimisation, a line to the minimum $\chi^{2}$ points (defined as lying within a $\Delta\chi^{2}$=2.3 above the minimal $\chi^{2}$), thus obtaining the linear relation between them. The fit is very good, $R^{2}=0.97$, reflecting the strong correlation, and from which the independent $1\sigma$ errors are derived (i.e., by the residuals around the minimal $\chi^2$ values). With this we obtain a best-fit (and $1\sigma$ errors) relative galaxies weight of $K_{gal}=11.4\pm0.6\%$, similar to the value expected based on our many previous HST analyses.

However, one cannot expect the power-law lumpy component to represent only the galaxies, nor the smooth component to represent solely the DM, so that trying to assess the true physical weight of each component would be unwarranted at present. One can only know for certain that the combination of the two with $K_{gal}$ as the relative weight, yields a good solution. It should also be mentioned, that we make a prior assumption on the range of sensible $K_{gal}$ values, so that the critical curves are not too smooth nor too lumpy. This is done by inspecting the resulting critical curves by eye, so that roughly, the degree of ``complexity'' of the critical curves is similar to that seen in the aforementioned HST-based analyses of some of the calibration-sample clusters, and in agreement with the general expectation for the (small) contribution of galaxies relative to the total mass.

\begin{figure*}
 \begin{center}
   \includegraphics[width=180mm]{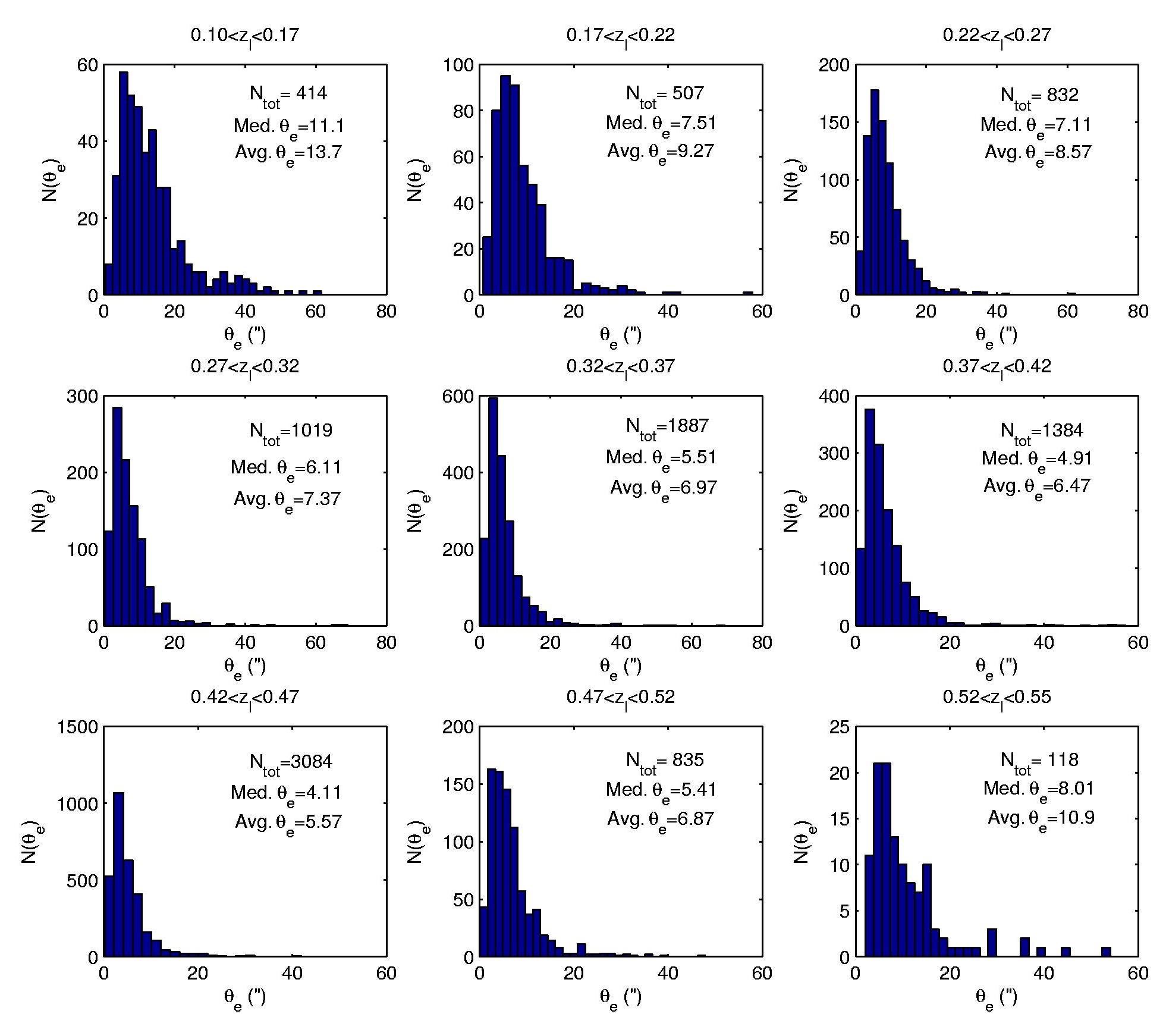}
 \end{center}
\caption{Einstein radius distribution ($z_{s}=2$) of $\simeq$10,000 SDSS clusters in different redshift bins. In each frame we specify the number of clusters in that redshift bin, and the median and mean radii of the distribution. The distributions are dominated, as expected, by galaxy and galaxy-group scale lenses, where the median and mean Einstein radii peak at $z_{l}\sim0.1-0.2$, although note that these rise again towards relatively higher redshifts of $\sim0.5$, so that large Einstein radii ($\theta_{e}>\sim40 \arcsec$) are more commonly deduced also for clusters at these redshifts (see also \S \ref{R_EinsteinRadius} and \ref{dependency}).}
\label{histAll}
\end{figure*}

For the normalisation factor, we obtain in the minimisation a best-fit value of $K_{q}=(51.6\pm1.9)\frac{d_{ls}}{d_{l}d_{s}}$ ($1\sigma$ errors). Accordingly, the $M/L$ related coefficient, $K$, equals $2.21 \times10^{28} / d_{l}^{~2-q}$, in units of [$gr/cm^{2-q}$; with $q$=1.2], from which we can deduce the explicit typical $M/L$ relation:

\begin{equation}\label{MLdeduced}
M_{(<\theta)}/L_{B}=8.7 \pm 0.3 \times10^{-5} L_{i}~\theta^{2-q}~~~ [M_{\odot}/L_{\odot}],
\end{equation}
where $q=1.2$ here, $\theta$ is in radians, and $L_{i}$ is the galaxy luminosity (in solar unit). For example, for a typical BCG as bright as $10^{10}~L_{\odot}$, this yields an $M/L_{B}$ value of $\sim120~[M_{\odot}/L_{\odot}]$ within $3\arcsec$, or e.g., $\sim180~[M_{\odot}/L_{\odot}]$ within $5\arcsec$. Note that this is not the typical $M/L$ per galaxy, but overall, a scaling which describes, per $L_{\odot}$, the total projected mass enclosed along the line-of-sight and within a cylinder of radius $\theta$ centred on a galaxy, and thus includes major contribution from the cluster DM halo along this line. Therefore, this $M/L$ term is coupled to the modelling procedure applied and include some internal rescalings and compensation to various effects such as the difference in the depth between the usual SDSS and HST imaging, and the red-sequence membership definition (\S \ref{results}), and are coupled to the LRG template and its possible minor redshift evolution (Ben\'itez et al. 2009). In fact, here we do not explicitly take into account this evolution of red-sequence galaxies and their host clusters, so that the resulting $M/L$ relation presented here may include a compensation to this effect, which although is expected to be minor (Ben\'itez et al. 2009), would be interesting to probe when a larger calibration sample is available.

With these best-fit values we analyse each of the SDSS reference-sample clusters sequentially in an automated way. The $1\sigma$ errors on these parameters mentioned above, propagate typically into a minor $\sim5\%$ error on the Einstein radius (of each individual cluster of the calibration sample), which may be too low in light of other uncertainties detailed in \S \ref{results}; accordingly, a more realistic error (or uncertainty) level is estimated as mentioned therein. The resulting Einstein radii of the SDSS blind analysis are compared to the results of detailed analyses in Figure \ref{TvTspec}, where a very good correlation with a small scatter is found ($R^{2}=0.97$, and deviation of less than $<5\%$). A more explicit example of the analysis results obtained by the different approaches is given in Figure \ref{comparison}.


\begin{table*}
\caption{Calibration-sample clusters (in arbitrary order). Columns are: cluster name; RA and DEC (J2000.0); photometric redshift and error from the Hao et al. (2010) catalog; Spectroscopic redshift; Einstein radii in arcseconds for $z_{s}=2$ adopted from detailed HST SL analyses, by our automated method in SDSS imaging, and by the Jackknife minimisation, respectively; Richness, number of galaxies assigned to the cluster in the Hao et al. catalog; reference for the adopted $\theta_{e}^{HST}$ value; other complementary references with relevant SL analyses. Spectroscopic redshift is assigned from the Hao et al. (2010) catalog if available, or else from the works mentioned below and references therein. Reference abbreviations are: \emph{Z}=Zitrin, A., unpublished; \emph{B05}=Broadhurst et al. 2005a;  \emph{HSP06}=Halkola, Seitz \& Panella 2006; \emph{L07/08/10}=Limousin et al. 2007/2008/2010; \emph{R09/10}=Richard et al. 2009/2010, and additional references therein; \emph{U09}=Umetsu et al. 2009; \emph{Z10/11a/b} = Zitrin et al. 2010/2011a/b; \emph{D11}=Donnarumma et al. 2011, and references therein; \emph{C12}=Coe et al. 2012. \emph{Comments:} We note that for RXJ2129 and A963, R10 obtained $\sim60-70\%$ lower values than we obtain with the same HST data, although large part of the difference may be as a result of a different Einstein radius definition. While we simply measure the area within the explicit critical curves and from that calculate the effective radius, they quote the radial distance in which the average $\kappa$ is equal to 1. Although both definitions are legit, they can only be expected to show similar values when the mass distribution is fairly round or symmetric. In addition, A2261 was previously analysed by us in Subaru imaging (U09) without any redshift or multiple-image information, and was recently thoroughly revised, as published in C12, based on new CLASH program imaging and combining several SL+WL techniques. According to the new photometric-redshift information for the multiple images found by our model (see C12), the Einstein radius seems to be about $\sim60\%$ lower of that reported initially in U09, and $\sim30\%$ lower than the value we had adopted here. We address the reader to these complementary works.}
\label{systemo}
\begin{tabular}{|c|c|c|c|c|c|c|c|c|c|c|c|}
\hline\hline
Identifier & RA & DEC & $z_{phot}$ & $z_{err}$ & $z_{spec}$ & $\theta_{e}^{HST}$ & $\theta_{e,auto}^{SDSS}$ & $\theta_{e,Jack}^{SDSS}$ & $N_{gal}$ & $\theta_{e}^{HST}$ ref & other refs\\
 &  \multicolumn{2}{c}{J2000.0}&  &  &  & arcsecs & arcsecs & arcsecs & && \\
\hline
   A1689 &  197.87295 & -1.3410050  & 0.2030 &   0.0180& 0.1832 & 46.0 &44.1& 38.2 &   142& Z & B05,HSP06,L07\\
   A1703 &  198.77197 &  51.817494  & 0.2690   & 0.0180 & 0.2800 & 28.0 &31.8 & 29.0  & 86& Z10 & L08,R09\\
   MS1358  & 209.96066 &  62.518110  & 0.3590 &0.0290 &  0.3273 & 13.0 & 13.4 & 12.0 &  69& Z11a & \\
   MACS1423 & 215.94948  & 24.078460  & 0.4410 &0.0950 & 0.5430 & 20.0 & 23.2 &20.7 &  16& Z11b & L10\\
   A1835 & 210.25886 &  2.8785320 &   0.2100 & 0.0390  & 0.2528 & 30.5& 30.8 &27.4& 65& R10 &\\
   Z2701 &  148.20456  & 51.885143  & 0.1920 & 0.0210   & 0.2151 & 9.0 &10.7 &8.8 &  11& R10 &\\
   A611  & 120.23668 &  36.056725  & 0.2900 & 0.0160  &  0.2873  &  21.0 & 21.5 &23.7&  59& R10 & D11 \\
   RXJ2129 & 322.41651 &  0.089227 &   0.2280 &0.0140 &   0.2339 & 21.1& 24.5 &23.9&  25& Z & R10 \\
   A963 &154.26499  & 39.047228 & 0.2230 &0.0110   & 0.2056  & 23.0 &23.9 &23.3 & 50& Z & R10\\
   A2261 &260.61326 &32.132572& 0.2250 & 0.0120  &  0.2233 & 35.0 & 36.3 &35.8 &  74& Z & U09, C12\\
\hline\hline
\end{tabular}
\end{table*}

\section{Results, Discussion, And Uncertainty}\label{results}

The sample analysed in this work is drawn from the Hao et al. (2010) SDSS cluster catalog. As mentioned in their work, Hao et al. (2010) have developed an efficient cluster finding algorithm named the Gaussian Mixture Brightest Cluster Galaxy (GMBCG) method. The algorithm uses the Error Corrected
Gaussian Mixture Model (ECGMM) algorithm (Hao et al. 2009) to identify the BCG plus red
sequence feature and convolves the identified red sequence galaxies with a spatial smoothing kernel
to measure the clustering strength of galaxies around BCGs. The technique was applied to the Data
Release 7 of Sloan Digital Sky Survey and produced a catalog of over 55,000 rich galaxy clusters in
the redshift range $0.1 < z < 0.55$. The catalog is approximately volume limited up
to redshift $z\sim0.4$ and shows high purity and completeness when tested against a mock catalog, and when compared to other well-established SDSS cluster catalogs such as MaxBCG (Koester et al. 2007; for more details see Hao et al. 2010).

We go over the Hao et al. (2010) catalog, and apply the method described above (\S \ref{model}) to each cluster, deriving its resulting Einstein radius and mass. We present here the results from the first 10,000 clusters analysed. In practice these 10,000 SDSS clusters comprise only a relatively small fraction ($\sim20\%$) of the full catalog coverage, whose analysis results we aim to presented in a future work, once a larger calibration sample is available.

\begin{figure}
 \begin{center}
   \includegraphics[width=90mm]{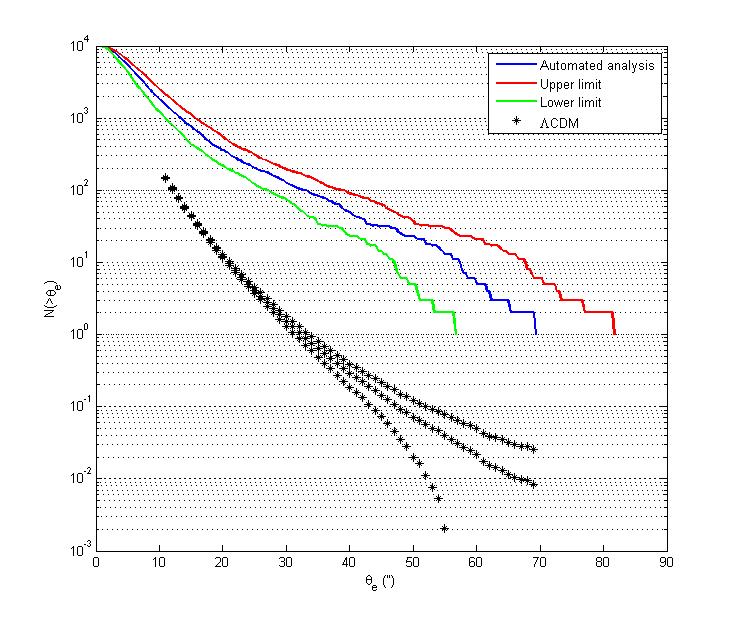}
 \end{center}
\caption{Cumulative Einstein radius distribution from $\simeq$10,000 SDSS clusters ($0.1<z<0.55$). The cumulative distribution, and its upper and lower $1\sigma$ limits, are shown in \emph{blue, red, and green solid lines}, respectively. Also plotted is the distribution predicted by the semi-analytic calculation of Oguri \& Blandford (2009), normalised to the effective sky area of our sample (\emph{black asterisks}, including errors). The two distributions disagree in two main aspects: there is a $\sim1-2$ orders-of-magnitude number discrepancy between them, and in addition, the two distributions have different slopes. The origin of the discrepancy is not clear and will be investigated elsewhere, although it may be as a result of different mass limit, and the choice of concentration-mass relation and mass function used in the semi analytic calculation. Correspondingly, we find a higher abundance of large $\theta_{e}$ clusters than predicted by the semi-analytic calculation. Our analysis yields $\sim40$ candidates with $\theta_{e}>40 \arcsec$ ($z_{s}=2$), with a maximum of  $\theta_{e}\simeq69\pm12\arcsec$ ($z_{s}=2$) for the most massive cluster. Interestingly, this value is in agreement with the estimate by Oguri \& Blandford (2000) for the largest Einstein radius. For more details see \S \ref{results}.}
\label{cumulate}
\end{figure}

\begin{figure}
 \begin{center}
   \includegraphics[width=90mm]{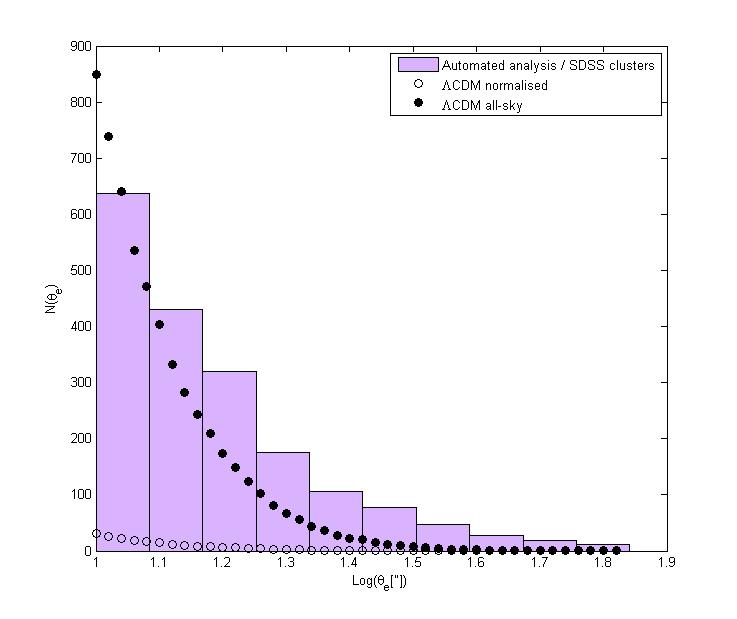}
 \end{center}
\caption{To assess the difference from the semi-analytic expectation by Oguri \& Blandford (2009; see also Figure \ref{cumulate}), we compare the width of the tails for $\theta_{e}>10\arcsec$, which is the lower limit taken in their calculation. The histogram shows the results from 1851 SDSS clusters following our analysis; the filled-circles curve shows the all-sky distribution from Oguri \& Blandford (2009), and the open-circles curve shows the same distribution normalised to the same sky area as our distribution. Both distributions are (semi) log-normal although with two main differences. The Oguri \& Blandford (2009) distribution has a width of $\sigma=0.1448$ (in $Log(\theta_{e})$), while our distribution shows a slower (or wider) decrease, with $\sigma=0.2436$. Also, the overall number of clusters in their analysis for the same sky area, is much lower. For more details see also Figure \ref{cumulate} and \S \ref{results}.}
\label{diffOguri}
\end{figure}


\subsection{Einstein Radius Distribution}\label{R_EinsteinRadius}
The resulting Einstein radius distribution for this sample is seen in Figure \ref{histAll} as a function of lens redshift (for constant $z_{s}=2$), along with the average and median Einstein radii for each redshift bin, which evolve in redshift and peak at $z_{l}\sim0.1-0.2$ (for $z_{s}=2$), as may be generally expected given the hierarchical growth history of clusters and the distances involved in lensing (this is further discussed in \S \ref{dependency}). The Hao et al. (2010) cluster catalog lists clusters with at least 8 members within 0.5 Mpc from the BCG. This low limit results in a realistic domination of galaxy-scale lenses ($\theta_{e}$ of an order of a few arcseconds), which are usually not massive enough to form impressive lenses with large Einstein radii and many multiple-images. The more interesting information may be the higher end of the distribution at larger radii. The concept of the largest Einstein radius in the Universe and the expected abundance of large lenses have been discussed thoroughly in the literature, and are especially of high interest as they teach us about the reliability of the standard $\Lambda$CDM model in predicting these extreme cases, as the $\Lambda$CDM model does not favor the formation of giant lenses (e.g., Broadhurst \& Barkana 2008, Sadeh \& Rephaeli 2008, Zitrin et al. 2009a).

Note that clusters with large Einstein radii are found also towards higher redshifts. In addition, though not included in this work, the largest known lens to date, MACS J0717.5+3745 is at a similarly high redshift of $z_{l}\simeq0.55$, with $\theta_{e}\simeq55\arcsec$  for $z_{s}\sim2.5$ (see Zitrin et al. 2009a). Due to a very shallow mass distribution in this cluster (Zitrin et al. 2009a), for $z_{s}=2$ the Einstein radius will be only slightly lower, around $\theta_{e}\sim50\arcsec$ (see also recent paper by Limousin et al. 2011 for new redshift information for this cluster). The exact number will be derived elsewhere, in the framework of the CLASH program. The abundance of larger lenses at these redshifts is caused usually (e.g., Zitrin et al. 2011a), by a spread-out, unrelaxed matter distribution. At these higher redshifts many clusters are not yet relaxed and still undergo mergers, so that the mass distribution is already sufficiently dense for significant lensing, but widely-distributed so that the Einstein radii of the different substructures are merged to form extended critical curves (e.g., Torri et al., 2004, Dalal, Holder, \& Hennawi 2004). On the other hand, at a lower redshift, more concentrated clusters are those yielding larger Einstein radii, as there is more mass in the centre enhancing the critical area (see also \S \ref{dependency}).

The blind analysis performed here yielded initially 69 candidates with $\theta_{e}>40 \arcsec$ ($z_{s}=2$), many coincident with various Abell or MACS clusters. We visually inspect each of these clusters and find that some are boosted by various effects detailed below (we omit these clusters from our further analysis), but infer that at least about half of these are most likely real giant-lens candidates, with a maximum of  $\theta_{e}\simeq69\pm12 \arcsec$ ($z_{s}=2$) for the most massive candidate. We direct the reader to works by Hennawi et al. (2007a) and Oguri \& Blandford (2009) which have investigated in detail the Einstein radius abundance on various scales, based on simulations and $\Lambda$CDM expectations, and taking into account triaxialities which induce a prominent lensing bias.

Our realistic, observationally-based results free from lensing bias, are compared to some such expectations explicitly in Figure \ref{cumulate}, where we plot the cumulative distribution of clusters above each radius with the expected $1\sigma$ errors, propagated from the errors on the best-fitting parameters as described in \S \ref{R_EinsteinRadius}. Note, the lower limit shifts the maximal Einstein radius from $\theta_{e}\simeq69 \arcsec$ to $\theta_{e}\sim57 \arcsec$ ($z_{s}=2$), close to that of the largest known lens, MACS J0717.5+3745 (Zitrin et al. 2009a). We note that Oguri \& Blandford (2009) who examined in detail the Einstein radius distribution based on semi-analytic expectations, have derived maximal Einstein radius values of $\sim60\arcsec$, but these as shown in their work are very susceptible to the cosmological parameters in general and to $\sigma_{8}$ in particular, and can reach (within the $3\sigma$ confidence) values that are nearly twice as high. Their expected distribution, scaled to the same sky area as our sample and with WMAP7 parameters (Komatsu et al. 2011), is overplotted in Figure \ref{cumulate}. Aside for an agreement between their expected largest Einstein radius and the largest lenses found in our analysis, the two cumulative distributions clearly disagree. Although normalised to the same effective sky area, there is a $\sim1$ order-of-magnitude number difference for small Einstein radii, which reaches a $\sim2$ orders-of-magnitude difference for higher Einstein radii, so that in addition, the two distributions have also different slopes. The origin of the discrepancy is not clear, but part of the difference may be due to a different (lower) mass limit probed by the two methods. In addition, the effect of the concentration-mass ($c-M$) relation and the chosen mass function used in semi analytic calculations should clearly have a strong influence on the resulting distribution (e.g., Duffy et al. 2008, Macci\'o et al. 2008, Prada et al. 2011; for differences among various $c-M$ relations), as higher concentrations entail higher inner mass and Einstein radius. We leave the examination of how these may influence the cumulative distribution, for future work.

To assess the difference from the semi-analytic expectation by Oguri \& Blandford (2009; see also Figure \ref{cumulate}), we compare the width of the tails for $\theta_{e}>10\arcsec$, which is the lower limit taken in their calculation. As can be seen in Figure \ref{diffOguri}, both distributions are (semi) log-normal although with two main differences. The Oguri \& Blandford (2009) distribution has a width of $\sigma=0.1448$ (in $Log(\theta_{e})$), while our distribution shows a slower (or wider) decrease, with $\sigma=0.2436$. This difference is commensurate with the different decline of the cumulative distribution seen in Figure \ref{cumulate}. Also, the overall number of clusters in their analysis, for the same sky area, is much lower, although this may be, as mentioned, entailed by the different mass limits probed by each method, which will be checked in future work. As mentioned above, one of the most interesting aspects of our study, namely the largest Einstein radius, is in good agreement with the estimation by Oguri \& Blandford (2009).

\subsection{Uncertainty and Error Estimation}\label{R_EinsteinRadius}

Various factors of error should be taken into account when addressing the results reported in this work, though these are mostly statistical and therefore due to the extensively large sample are not significant. In addition, these uncertainties arise mainly from the data themselves, so that applying the method presented here to higher-end, dedicated cluster-survey data (e.g., the expected J-PAS survey; Moles et al. 2010), should produce much cleaner results.

\subsubsection{Possible Factors of Error}\label{factorsoferror}
The first error factor we have investigated is the lens photometric-redshift error. The typical photo-$z$ uncertainty for the sample BCGs (by which we determine the lens distance) is 0.015. A $\sim10\%$ error in the lens redshift can be translated into a noticeable ($>10\%$) difference in the measured Einstein radius, and only about half of the sample has spectroscopic redshifts for the BCG, thus the results for some of the clusters are affected by this error. As can be seen in Hao et al. (2010), the photometric redshifts were tested against the spectroscopic redshifts where possible, yielding a very tight relation strengthening the confidence in them. We have tested the effect of the photometric redshifts on the calibration sample, and regenerated Fig. \ref{TvTspec} based on the photometric redshifts (instead of the spectroscopic redshifts; see also Table \ref{systemo}). Only slight differences are seen and the overall scatter remains essentially the same. To assess the effect of the photo-$z$ error more quantitatively, we analysed a sample of 500 random clusters (detailed in \S \ref{jack}) with the catalog photometric redshifts, and then repeated the analysis by photometric redshifts drawn randomly from a normal distribution centred on the catalog photometric redshift for each cluster, with a width of $\sigma=0.015$ (which is the photo-$z$ error quoted in Hao et al. 2010). From this we indeed obtain a low uncertainty of only $1.15\%$ on the cumulative Einstein radius distribution, and for the differential (log-normal) distribution, differences of only $0.38\%$ and $0.92\%$ on $\langle Log(\theta_{e}) \rangle$ and $\sigma$, respectively.

\begin{figure}
 \begin{center}
   \includegraphics[width=90mm]{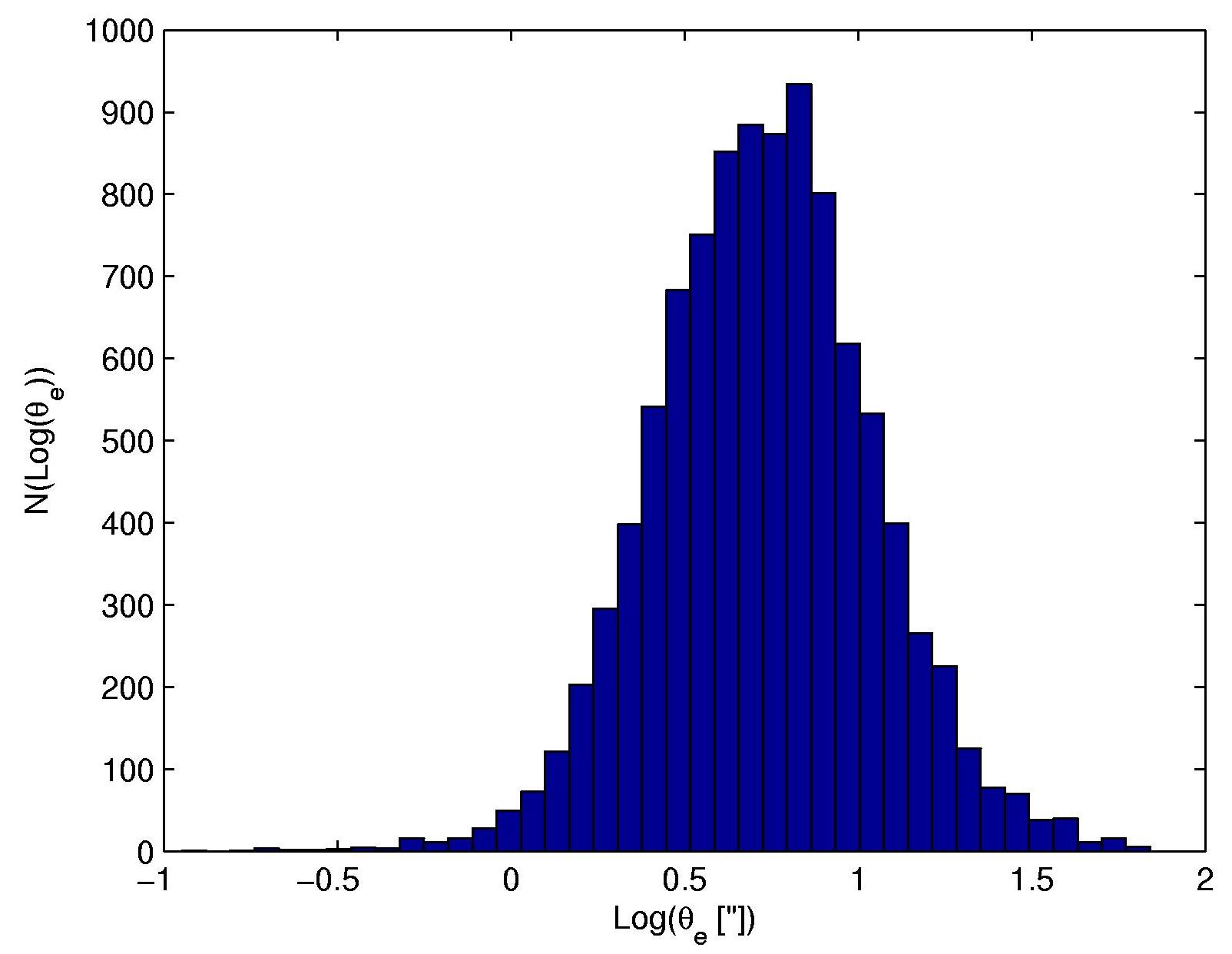}
 \end{center}
\caption{Einstein radius (log) distribution, from $\simeq$10,000 SDSS clusters ($0.1<z<0.55$). The sample has a log-normal distribution, with $\langle Log(\theta_{e}\arcsec)\rangle=0.73^{+0.02}_{-0.03}$ and $\sigma=0.316^{+0.004}_{-0.002}$.}
\label{reHistLog}
\end{figure}

\begin{figure}
 \begin{center}
   \includegraphics[width=90mm]{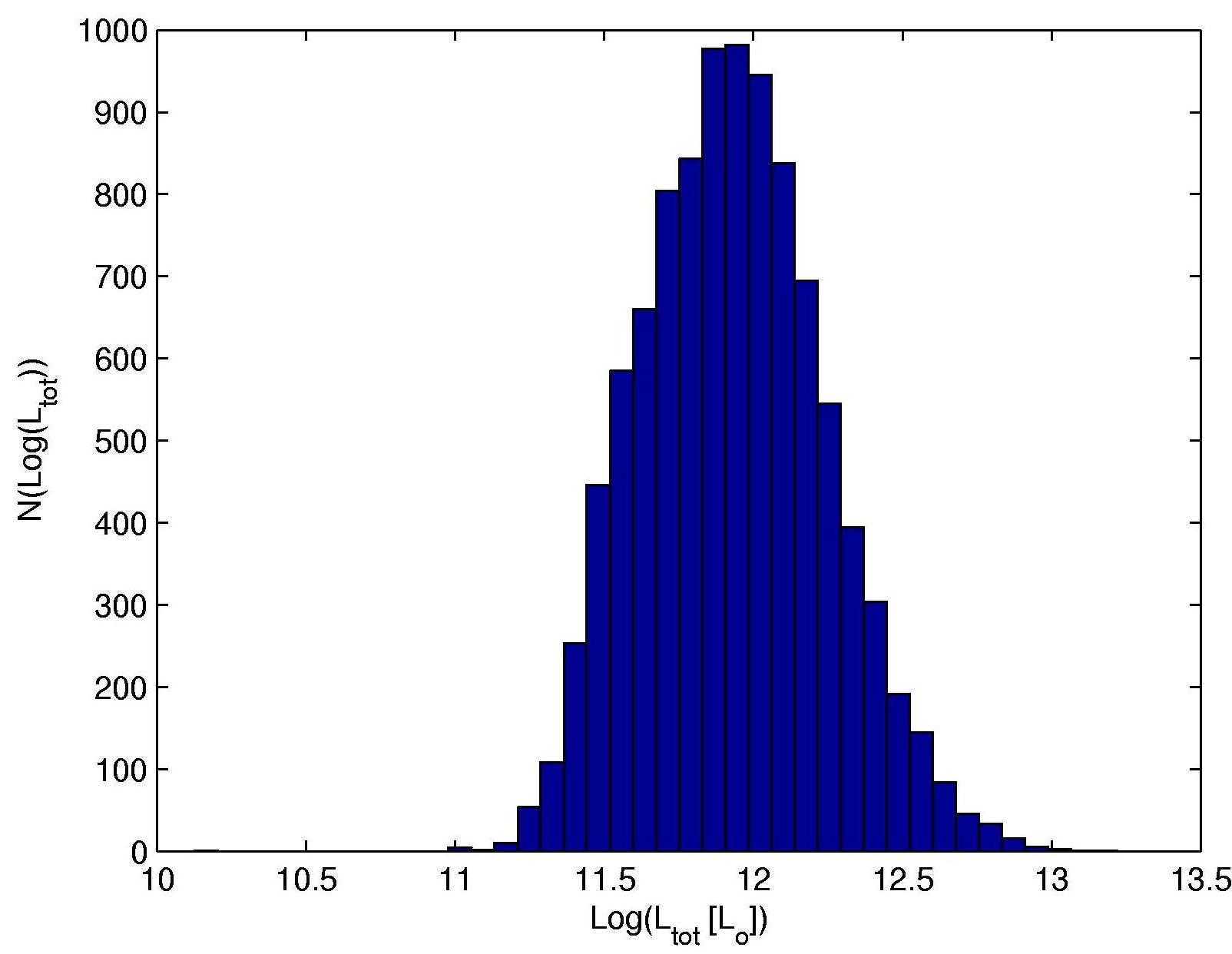}
 \end{center}
\caption{Luminosity distribution (log). Plotted is the histogram of total B-band solar luminosities for each cluster, i.e., the sum of all cluster member luminosities. We converted the SDSS $r$-band luminosities to B-band (Vega) luminosities by the LRG template characterised in Ben\'itez et al. (2009).}
\label{lumHistLog}
\end{figure}
Second, the SDSS imaging is shallower than typical HST imaging dedicated to SL analysis. Correspondingly, and supplemented by the conservative cluster-finding algorithm, some of the cluster members are overlooked and often not associated with the cluster, and only the brighter galaxies are incorporated. Luckily, these are also the more massive galaxies and thus the effect on the lens model is minor. In addition, the inclusion of (less-massive) cluster members is known to affect locally the shape of the critical curve, but not to change their overall size (e.g., Flores, Maller \& Primack 2000, Meneghetti et al. 2000). The constant $K_{q}$ which was iterated for (which includes the $M/L$ ratio) is probably boosted by the relative loss of galaxy mass-representations in our modelling of the SDSS catalog. This, however, can be very well assumed to be a relatively constant ratio and thus, along with the $M/L$ intrinsic scatter, does not affect substantially the results, as can be seen in the calibration sample comparison, where a clear consistency is found. This may not at all be surprising, since the modelling
here is based on simple physical considerations: it has
been well established that light approximately traces mass,
and clearly, a reasonable $M/L$ relation can be incorporated.

Another factor of possible contamination is the lower resolution of SDSS images compared with typical lensing images by HST. This we find may result in local overestimation of the BCG, if another cluster member is found too close to the BCG core to be resolved, thus boosting the Einstein radius (and mass), especially for higher-redshift lenses. The reason is that the smoothing procedure, or the (2D polynomial) fit, is dominated by the BCG. Therefore, although the lumpy (galaxy) component in such scenarios should not have a substantial effect, the smooth component will be over-boosted in the middle (since the BCG would be too bright), thus pushing outwards the Einstein radius. However, this chance alignment is naturally not too common, and in any case will affect mostly the lower end of the distribution, i.e., clusters with small Einstein radius that is dominated fully by the BCG. Clusters with large Einstein radii will be less susceptible to such contamination as the critical curves are not fully dominated by the BCG and contain substantial contribution from other massive cluster members as well.

A known factor of systematic error considered in related work on various samples of (often SDSS) clusters, is the miss-centring of mass with respect to the BCG (e.g., Becker et al. 2007, Johnston et al. 2007a, Rozo et al. 2009,2010, Oguri \& Takada 2011).  In that sense, methods which depend on a predefined centre may be affected from a scatter in the location of the BCG with respect to the true centre-of-mass, if the prior is constantly assumed to lie at the very centre of the cluster. However, in our method, there is no need to predefine the exact centre-of-mass. The smoothing procedure we implement has the advantage of being independent from a predefined centre, and the result is ultimately determined simply and directly by the galaxy (light) distribution. In fact, this has enabled us to find various such shifts between the BCG and the centre of (dark) mass (e.g., Zitrin et al. 2009b, Umetsu et al. 2011a).

However, if the GMBCG catalog itself has misidentified a galaxy as the BCG, which we use as the centre-of-frame for our analysis, this may entail a shift in the analysed field, so that in principle some relevant galaxies may lie outside it. Nevertheless, since the Hao et al. (2010) catalog considers galaxies within 0.5 Mpc, even for the highest redshift clusters of the sample ($z=0.55$), this size translates into $\simeq78$ arcseconds. Since the Einstein radius is determined by the mass enclosed within it, and since only less than a handful of clusters may have such a large Einstein radius (following the $1\sigma$ upper limit, see Figure \ref{cumulate}), this may only have a negligible effect over the whole sample.

It should be noted that the results for $z_{l}>0.43$ should be more cautiously addressed, as the catalog is officially volume limited up to this redshift due to the luminosity cuts that require potential member galaxies to be brighter than 0.4L*, where L* is the characteristic luminosity in the Schechter luminosity function. Also, for higher redshifts, the different red-sequence criteria ($r-i$ instead of $g-r$, see Hao et al. 2010) may come in play and input some more noise, mostly with respect to the richness level, so that overall one should expect fewer members assigned for $z_{l}>0.43$ clusters relative to clusters below this redshift. In order to test this effect we repeated our analysis including only clusters in the volume limit of $z_{l}<0.43$ and verified that only negligible differences are seen with regard to the Einstein radius distribution (e.g., such analysis yields a log-normal Einstein radius distribution with $\langle Log(\theta_{e}\arcsec)\rangle=0.75$ and $\sigma=0.31$, similar to the full sample; see Figure \ref{reHistLog}). In addition, high redshift clusters in the calibration sample also show a satisfying result following the same scaling relation as lower redshift clusters (with a scatter of up to $\simeq15\%$ with the best-fitting parameters, or up to $\sim5\%$ scatter with the Jackknife minimisation, see \S \ref{jack}). Also, if we exclude these from the calibration-sample minimisation, the best-fitting parameters differ by less than $3\%$ from those obtained with the full sample.

\begin{figure}
 \begin{center}
   \includegraphics[width=90mm]{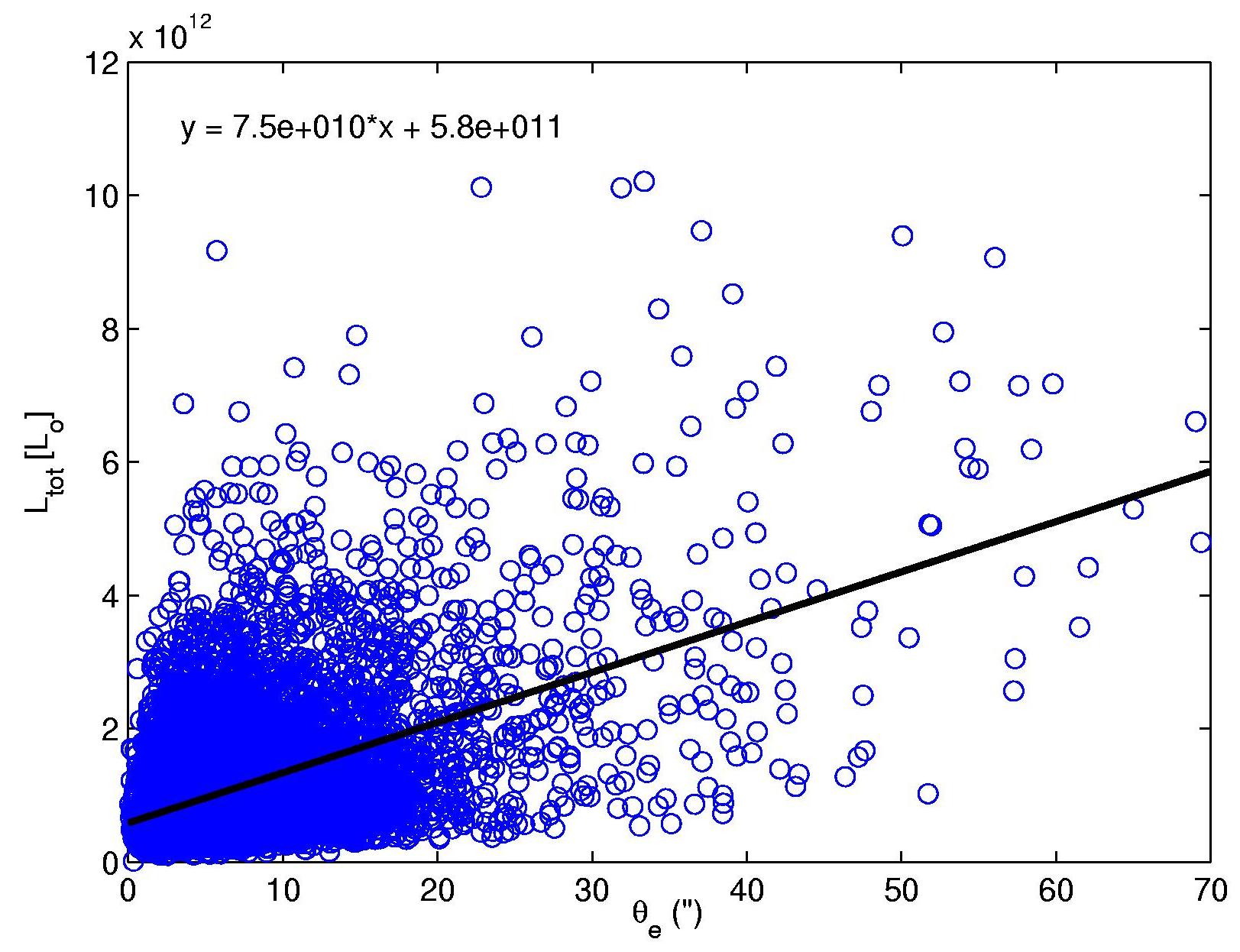}
 \end{center}
\caption{Total luminosity versus Einstein radius. As is evident, the total luminosity itself is not an accurate indicator of the Einstein radius. Following the more realistic procedure described in this work is necessary in order to obtain a reliable mass model and consequently a reliable Einstein radius distribution.}
\label{lum_re}
\end{figure}

\begin{figure}
 \begin{center}
   \includegraphics[width=90mm]{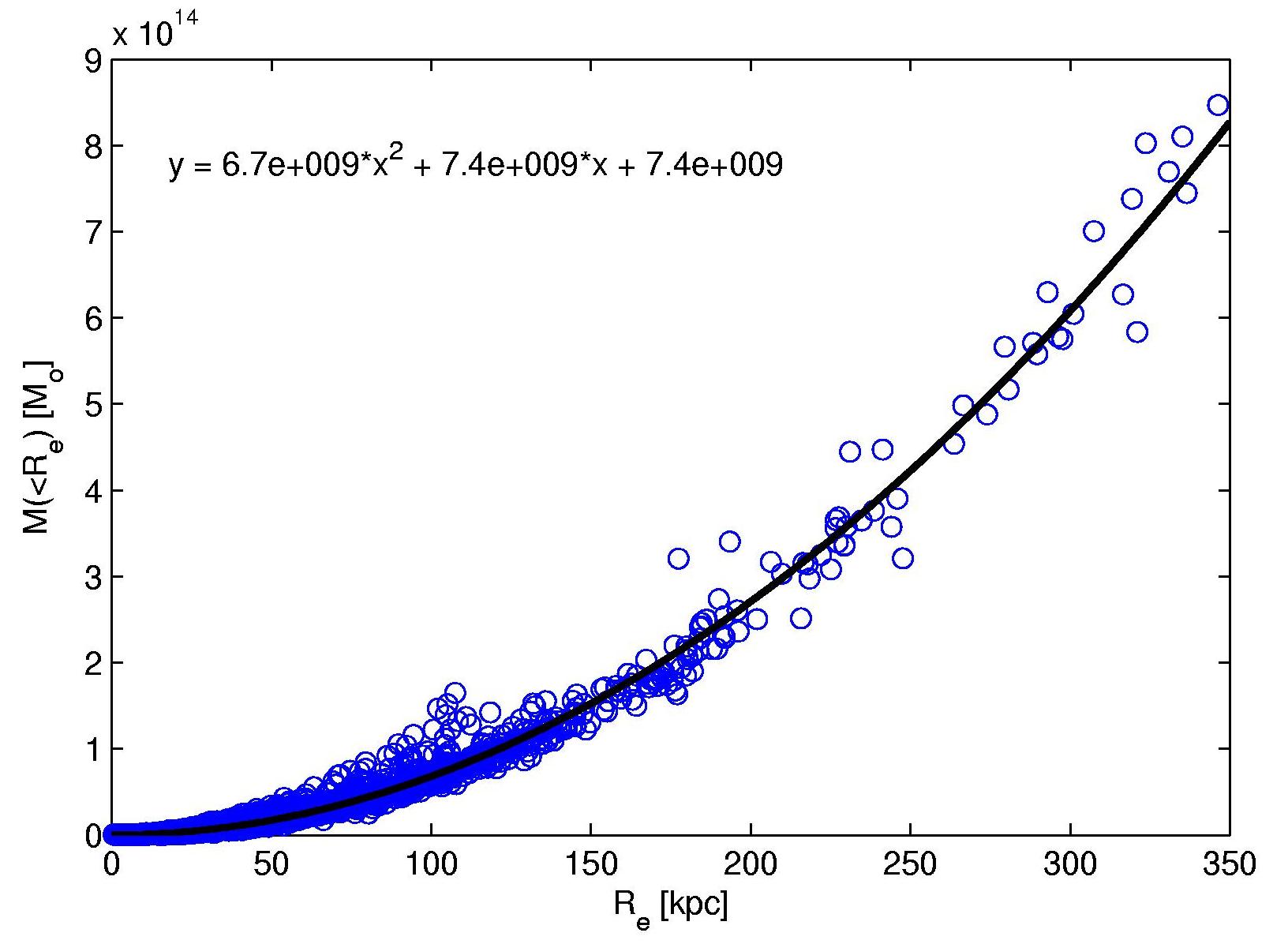}
 \end{center}
\caption{Consistency check for our procedure. The Einstein mass versus the Einstein radius of the $\simeq$10,000 clusters analysed here. As can be expected, the Einstein radius and mass correlate with a square relation, whereas some scatter is seen since naturally the lenses are not strictly symmetric. Spurious detections did not follow the presented relation, aiding us in excluding them from further analysis.}
\label{mre}
\end{figure}

We have identified within the 69 initial candidates with $\theta_{e}>40 \arcsec$ ($z_{s}=2$), several clusters that were misidentified as higher redshift clusters (according to their observed BCG), though they are most likely substructures of a foreground more massive (and known) cluster on the same line-of-sight. This boosts significantly the Einstein radius, and such cases as mentioned were omitted from our further analysis. Due to the low chances of such alignments and resulting misidentifications, the effect of this on the full sample and especially on the lower $\theta_{e}$ regime, is expected to be minimal.

\subsubsection{Quantification of Errors and Uncertainty}\label{jack}
In order to assess better the amount of statistical uncertainty caused by the various factors (e.g., \S \ref{factorsoferror}), we have first examined by eye a sample of 100 random clusters from the catalog and the critical curves generated for them by our automated modelling. We found only 3 clusters whose Einstein radius is boosted due to an unresolved galaxy near the BCG, and $\sim$15 more clusters with some galaxies that by eye do not necessarily seem to have similar colors, so that if these are misidentified as cluster members, may be introducing some additional noise. Clearly, this designation is not purely objective, but still allows us to conclude, in addition to the other consistency checks we performed, that the overall noise level in our analysis is reasonable. As an additional complementary step, and regardless of the calibration sample, we searched for SDSS $z_{s}\sim2$ arcs found in Bayliss et al. (2011b) and examined how well our critical curves could in principle reproduce these (giant) arcs. Although the location of the arcs is not used as input, our blind analysis automatically reproduces critical curves that pass through them as expected, strengthening further our automated approach. An example is given in Figure \ref{Bayliss}.

To explicitly quantify the errors and level of uncertainty we perform two majors procedures. Firstly, we perform a ``Jackknife'' minimisation: on top of the $\chi^2$ (eq. \ref{chi2}) minimisation with the full calibration sample (to find the best parameters for the blind analysis), we perform the minimisation 10 more times, each time omitting one cluster from the fit, and then analysing it with our automated procedure to examine how well the Einstein radius is estimated. We note, that by doing so the best-fit values for $K_{gal}$ and $K_{q}$ in each such iteration distribute around the best-fitting parameters when minimised by all ten clusters together, with values up to $\simeq3\sigma$ away. With this, we obtain that the Einstein radii for all ten clusters are estimated within $\sim17\%$ of their reference value (according to HST-based detailed analyses with multiple-images as input), while 9 clusters show a scatter of up to $\sim13\%$. These (only) represent how well each of the Einstein radii of the calibration sample can be reproduced individually.

Therefore, secondly, we wish to examine how the ($1\sigma$) errors drawn from the full reference sample minimisation are propagated into the statistical results, since these should depend on other quantities such as, e.g., the number of lenses per (Einstein radius) bin. For that purpose, we analyse 500 random clusters with the best-fitting parameter values, and then repeat the analysis marginalising over the $1\sigma$ errors. For the differential, log-normal Einstein radius distribution, these result in differences of $^{+2.8\%}_{-3.5\%}$ on $\langle Log(\theta_e)\rangle$, and $^{+1.31\%}_{-0.7\%}$ on $\sigma$, and upper and lower limits of $\sim18\%$ on the cumulative Einstein radius distribution (Fig. \ref{cumulate}). We take these to represent the level of (statistical) uncertainty in our analysis.

It should be mentioned, that although the sample analysed here was not selected based on mass or arc abundance and thus is not biased in terms of lensing, the calibration sample used to determine the model parameters ($K_{q}$ in particular) consists of 10 well-known massive clusters, which might introduce a systematic error boosting the Einstein radii. Though low and moderate-mass lensing clusters are hard to model for comparison due to lack of multiple-image constraints, the calibration sample contains clusters with as few as 11 members, and as many as 142 members, thus spanning nearly the full richness range of the probed SDSS sample. The possible bias might be further looked into by comparing galaxy and group-scale lenses with known prominent arcs, often found in systematic surveys for gravitational arcs (e.g., Sand et al. 2005, Hennawi et al. 2008, Kubo et al. 2010, Bayliss et al. 2011a,b, Wen, Han \& Jiang 2011), which should also be useful for extending the calibration sample and examining further this effect.

We note that due to the approach implemented here which does not use multiple-images as input, the profiles and magnifications are not well constrained for each cluster, and the only relevant measure which we refer to is the effective Einstein radius (and enclosed mass), as seen in Figure \ref{QSmag}. Naively, one could in principle derive the mass profile for each lens by simply assuming different fiducial source redshifts and calculate the enclosed mass by implementing their distance-redshift relation, but this would be premature at this stage, as though the parameters maintained constant here (on typical values) do not considerably affect the critical curves shape and size, the mass profile is susceptible to these and thus a separate calibration is required for each source redshift based on the full reference sample. This however may indeed be plausible, as we intend to probe in future work.

\subsubsection{Consistency checks}
To further verify that the data used here is reasonable for our purpose, especially the luminosity of clusters members to which our method is coupled, we perform a few simple self-consistency checks. The overall Einstein radius distribution is plotted is Figure \ref{reHistLog}, and is clearly log-normal in shape, with $\langle Log(\theta_{e}\arcsec)\rangle=0.73^{+0.02}_{-0.03}$ and $\sigma=0.316^{+0.004}_{-0.002}$. The luminosity distribution is plotted in Figure \ref{lumHistLog} for comparison, and for each cluster we explicitly compare in Figure \ref{lum_re} the total luminosity to its resulting Einstein radius, where it is importantly evident that the total luminosity itself is not an accurate enough measure of the Einstein radius. Following a more realistic procedure as described in this work is necessary in order to obtain a reliable mass and Einstein radius distribution. More explicitly, as the mass is more concentrated than the light, one must choose a more concentrated representation for the galaxies (e.g., the power-law used here), which is then simply scaled by the luminosity. Similarly, we stressed that the DM is well represented by smoothing the galaxies mass distribution, which is more efficient in practice for the inner SL region, than e.g. assuming a symmetric mass distribution such as NFW which often does not allow to immediately uncover the multiple-images by the model.

In Figure \ref{mre} we plot the enclosed mass versus the Einstein radius for each cluster. Although a tight relation is expected directly from the lensing equations, this constitutes an important self-consistency check. We simply measure the Einstein radius for each cluster as the area enclosed within the critical curves (exploiting the magnification sign-changes to estimate this automatically), where the mass is measured by summing the surface-density in the pixels which fall within the critical curves. The Einstein masses correlate well with the Einstein radii, with a square relation as expected, and with a reasonable scatter since the clusters cannot be expected to be strictly symmetric. Explicitly, the $R^{2}$ of the fit is 0.985, and the mean scatter is lower than $20\%$ (although there is an excess in the scatter of up to almost $100\%$ around 100 kpc for some individual clusters, probably due to the different factors of error elaborated in \S \ref{factorsoferror}). Also, in this consistency check, clusters spuriously assigned with large Einstein radii due to effects detailed above did not follow the expected relation, aiding us to exclude them from our further analysis.

\subsection{Correlations with cluster parameters}\label{dependency}

Since the Einstein radius correlates with the mass interior to it, some dependence on the examined cluster parameters which are related to the observed mass, such as redshift, richness, and luminosity, can be expected. For example, in Figure \ref{histAll} we showed the Einstein radius distribution in different redshift bins, where it is evident that large Einstein radii are observed more frequently in the lower ($z\sim0.1$) and higher ($z\sim0.5$) redshifts of the sample, whereas in intermediate redshifts the mean Einstein radius is smaller.

\begin{figure}
 \begin{center}
   \includegraphics[width=80mm]{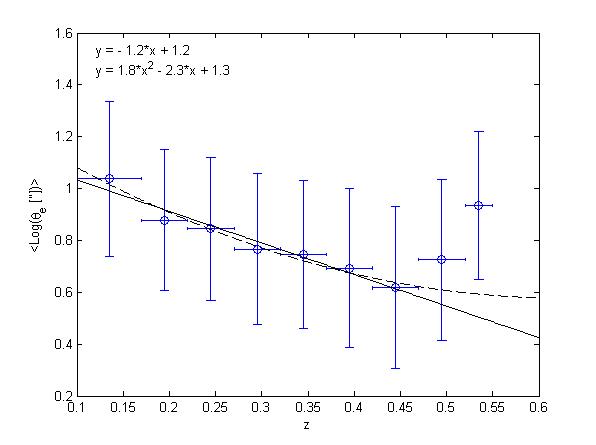}
   \includegraphics[width=80mm]{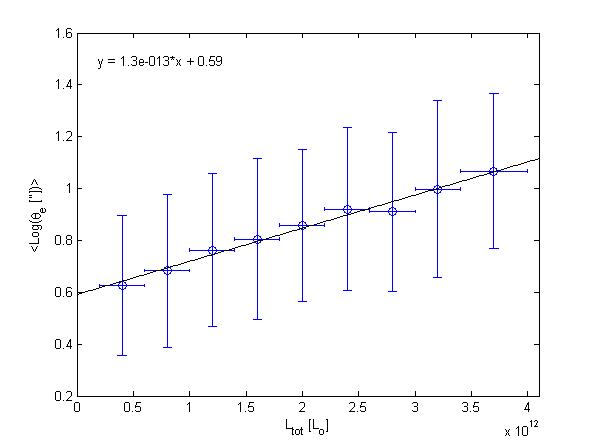}
   \includegraphics[width=80mm]{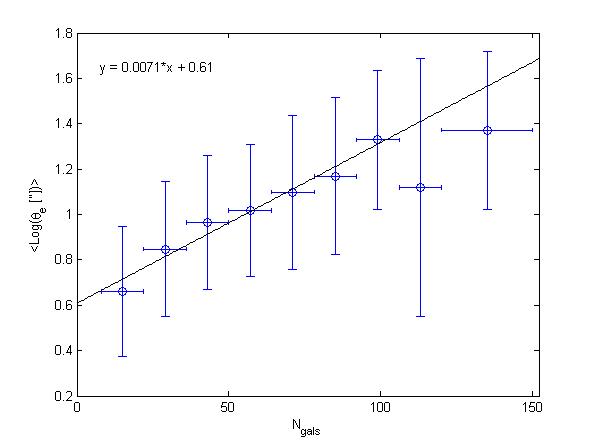}
 \end{center}
\caption{Evolution of the mean effective logarithmic Einstein radii with redshift (\emph{top}), total luminosity (\emph{centre}), and richness (\emph{bottom}). The horizontal error bars mark the bin widths, and the vertical error bars mark the width of the distribution in the corresponding bin, $\sigma$. In each plot we least-square fit a linear curve to the data (solid lines), where for the redshift plot (top) we also show a second-order fit (dashed line). The curve fits in the redshift and richness plots include only the first seven bins, due to incompleteness of the catalog at higher redshifts, governed by higher richness and luminosity clusters. In the total luminosity plot, we do not show the full luminosity range, since there are too few clusters to deduce a representative distribution for higher luminosity bins than those shown. See \S \ref{dependency} for more details.}
\label{3params}
\end{figure}

\begin{figure*}
 \begin{center}
   \includegraphics[width=140mm,trim=0mm 0mm 0mm 0mm,clip]{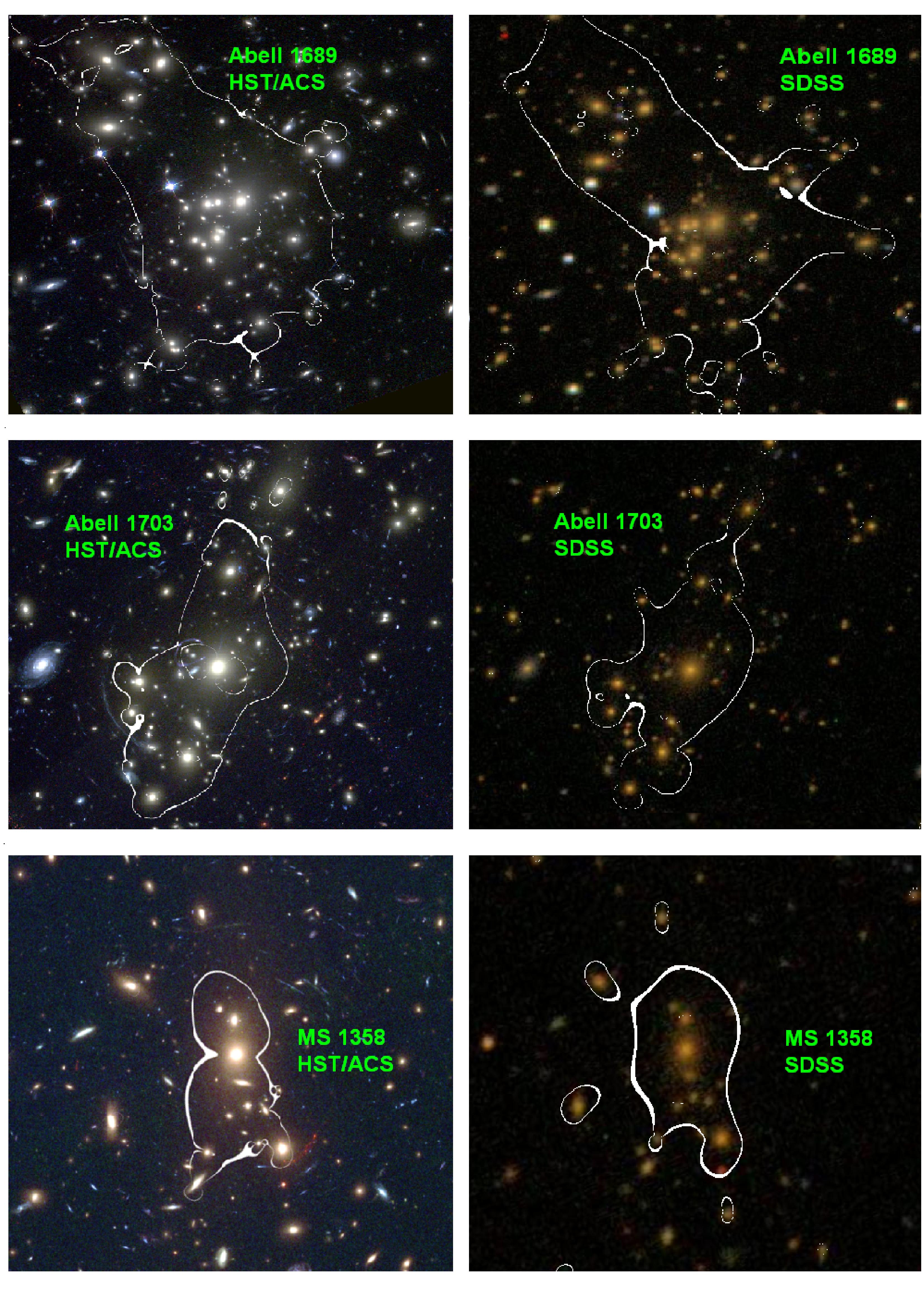}
 \end{center}
\caption{Comparison of three clusters from our reference sample covered by both HST and SDSS. The \emph{left} column shows for each cluster the resulting critical curves ($z_{s}=2$) of a detailed SL analysis in HST/ACS images and based on the incorporation of many multiple-images to fine-tune the mass model (e.g., Broadhurst et al. 2005a, Zitrin et al. 2010,2011b). The \emph{right} column shows the results of the ``blind'' automated method described here, on the same clusters in SDSS data, based on the light distribution of bright cluster members and without any multiple-images information. As can be seen, the automatic SDSS analysis produces overall similar critical curves to those derived by the detailed previous analyses. More importantly, though some local difference is seen between the two methods, the overall Einstein radius remains the same, as seen also in Figure \ref{TvTspec} for the full reference sample.}
\label{comparison}
\end{figure*}

\begin{figure*}
 \begin{center}
   \includegraphics[width=160mm,trim=-10mm 0mm 0mm 0mm,clip]{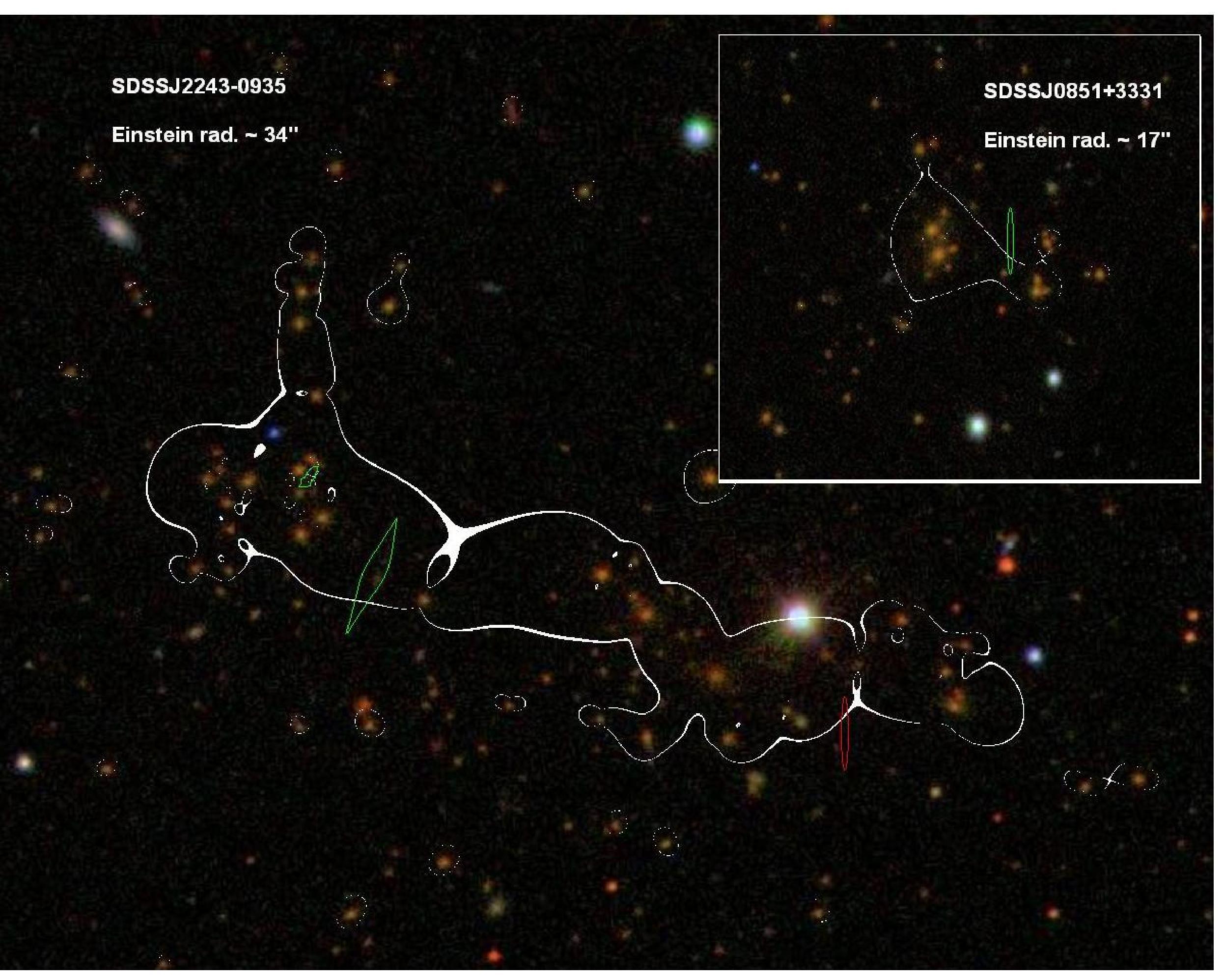}
 \end{center}
\caption{Comparison of two SDSS clusters with giant arcs from Bayliss et al. (2011a,b). The green ellipses mark the (approximate) locations of $z_{s}\sim2$ giant arcs identified by Bayliss et al. (2011b) in deep Gemini data, and the arc marked with a red ellipse is seen in available HST data, and may be at a slightly different redshift. Note that without using multiple-images location as input, the critical curves from our ``blind'' analysis pass through the giant arc as expected, strengthening further our automated approach. }
\label{Bayliss}
\end{figure*}

To further quantify this effect, and since Figure \ref{reHistLog} shows that the Einstein-radius distribution is log-normal, we plot the mean (and width) of the log-normal effective Einstein radius distribution in different redshift bins. The result is seen in Figure \ref{3params} (\emph{top}), where we also fit first- and second-order polynomials to the data. Although this tendency is only of the order of the log-normal distribution widths, the mean effective Einstein radii steadily decrease from $z\sim0.1$ to $z\sim0.45$, and then increase again (see Figure \ref{3params}). This tentative decrease of the mean effective Einstein radius with redshift may be related to cluster evolution. For example, lower redshift clusters, which have had more time to collapse, relax, and virialise, are expected to have more concentrated mass distributions and thus be stronger lenses (e.g., Giocoli et al. 2011). On the other hand, the tentative increase of the mean effective Einstein radii towards $z\sim0.5$ may be related to more substructured mass distributions, whose critical curves for the several merging subclumps are merged together to a bigger critical curve (e.g., Torri et al., 2004, Dalal, Holder, \& Hennawi 2004), although it is unclear at present how prominent is this effect.

It should be noted, however, that the cluster catalog is volume limited only up to $z\sim0.43$ where the tentative rise with redshift towards larger Einstein radii sets in. No weight should therefore currently be given to the two highest redshift bins since they may be affected by different selection criteria applied for cluster detection above $z_{l}\sim0.43$. To test this, we examined plots of the cluster luminosity and richness versus redshift, as larger $\theta_{e}$ clusters should have in principle higher luminosity and richness. A pronounced step is seen at $z=0.43$, so that the majority of luminous clusters ($\gtrsim2\times10^{12}~L_{\odot}$) is found at $z_{l}>0.43$, which most likely renders the rise in Einstein radii above $z_{l}>0.43$ a result of the different red-sequence criteria applied for higher-$z$ clusters. Given this, we ignore the mean effective Einstein radii above $z=0.43$, and concentrate on the indication of a decrease from low redshift towards $z=0.43$.

One should quantify the effect of geometry on the observed evolution of the mean (log) Einstein radius with redshift. The motivation is to check what is the contribution of pure geometrical effects, versus, say, evolutionary processes of the clusters. However, this can only be done qualitatively, since in order to know the geometrical dependence of the Einstein radius, one has to know the mass profiles in advance. For example, for a point mass $\theta_{e}\propto (d_{ls}d_{l}/d_{s})^{0.5}$, while for an isothermal sphere $\theta_{e}\propto (d_{ls}/d_{s})$. Generally, for a power-law projected mass distribution $\propto\theta^{-w}$, the angular Einstein radius scales with the distances as $\theta_{e}\propto (d_{ls}d_{l}/d_{s})^{1/w}d_{l}^{-1}$. If we take the power law to be isothermal, $w=1$, the angular Einstein radius (for constant $z_{s}=2$) decreases by $\sim25\%$ from the $z_{l}\sim0.13$ bin to the $z_{l}\sim0.45$ bin. For mass profiles steeper than isothermal, $\theta_e$ decreases more rapidly with lens redshift, while for flatter profiles it increases (or shows a more complex behavior such as increasing and then decreasing). The observed monotonic decline between $z_{l}\sim0.13$ and $z_{l}\sim0.45$ seen in Figure \ref{3params} is of $\sim40\%$, so that a steeper profile than isothermal ($w\simeq1.5$) is needed to fully explain this decline by geometrical means. Any possible significance of the redshift evolution of $\theta_e$ for cluster evolution can thus only be assessed once the mass profile is known.

We also compare the observed decline to semi-analytic calculations, in which standard concentration-mass relations and mass functions are incorporated. Quite independent of their detailed assumptions, such calculations yield \emph{increasing} mean Einstein radii with lens redshift, opposite of our result. One such semi-analytic calculation we use for comparison here (M. Redlich, private communication) is based on the Press \& Schechter (1976) mass function. Effective Einstein radii are derived from assumed relaxed (i.e.~non-merging), triaxial NFW haloes, adopting the concentration-mass relation from Jing \& Suto (2002) and including only halos with $M\ge10^{14}~M_{\odot}$. This calculation yields, similar to the result by Oguri \& Blandford (2009), a larger fraction of smaller Einstein radii than our findings, and the mean effective (logarithmic) Einstein radius increases with redshift, contrary to the decline we observe here. This however, may be partly explained (like the discrepancy in Figure \ref{cumulate}) by the lower-mass limit (more lower-mass clusters in higher redshifts would alleviate the discrepancy), or the choice of mass function and concentration-mass relation implemented therein. A more extensive calibration sample needs to be obtained before strict conclusions can be drawn. In principle, however, our results may help to pin down the true concentration-mass relation, in order to be compared with the evolution trends obtained in semi-analytic calculations, independent cluster evolution studies (e.g., Maughan et al. 2008 and references therein), or related numerical simulations (e.g. Duffy et al. 2008, Macci\'o et al. 2008, Prada et al. 2011).

Finally, we note that the (logarithmic) mean effective physical Einstein radius, i.e.~the angular Einstein radius times the angular-diameter distance to the cluster, is constant across the redshift range $0.1<z_{l}<0.43$ with $\langle Log(\theta_{e}~[$kpc$])\rangle=1.25\pm0.03$. Tentatively, near-constant physical Einstein radii with redshift can be achieved with standard, e.g.~NFW density profiles if the concentration-mass relation evolves steeply with redshift. Adopting a relation of the form $c\propto M^{-\alpha}(1+z)^{-\beta}$ (e.g.~Duffy et al. 2008), we find that $\beta\approx-3$ is required to reproduce the trend seen in Figure \ref{3params}. Current well-established concentration-mass relations extracted from numerical simulations find $\beta\sim-1$ (e.g., Bullock et al. 2001, Duffy et al. 2008). We emphasize once more that these conclusions are tentative and preliminary, as the results of this work should be revised once a more significant calibration sample is available. A more profound investigation of the indicated redshift evolution and its comparison to numerical simulations would thus be premature.

Apart for the redshift dependence, we repeated the above procedure and examined the mean effective logarithmic Einstein radii also in different total luminosity and richness bins (namely, how many red sequence galaxies are assigned to each cluster by the Hao et al. 2010 catalog). The evolution of $\langle Log(\theta_{e}\arcsec)\rangle$ and $\sigma$ for these is plotted in Figure \ref{3params} (\emph{centre} and \emph{bottom}), respectively. A (mild, given the distribution widths) trend is uncovered as a function of both luminosity and richness, so that on average, higher luminosity and richness clusters, show larger Einstein radii. These trends are quite expected, since richer clusters are naturally more luminous and more massive, correspondingly (for established mass-richness relations see, e.g., Rozo et al. 2009, Bauer et al. 2012). In addition, although not specifically shown here, for completeness we also examined both the richness and luminosity in the difference redshifts bins. We note that the richness is $\sim$constant throughout the volume-limited redshift range, while $\langle Log(L_{tot}~[L_{\odot}])\rangle$ monotonically increases by $\simeq0.4$ from $z\sim0.1$ to $z\sim0.45$, as can be generally expected from passive evolution of the cluster galaxies (although also here the trend is insignificant given the widths of the distribution in each bin, which is of the same size as the increase throughout the range, $\sigma\simeq0.4$).

We note, in addition, that with respect to the \emph{widths} of the logarithmic distributions in different redshifts, luminosity or richness bins, we do not uncover any prominent trend (which could withhold information on, say, the level of different population mix in the different bins).

\section{Summary}

In this paper we presented an automated strong-lens modelling tool, which is used to efficiently derive the Einstein radius and mass distributions of 10,000 SDSS clusters, found by the Hao et al. (2010) cluster-finding algorithm in DR7 data. We adopt the well-tested approach that light overall traces mass (with a more realistic representation of the galaxies and DM, see \S \ref{model}), and normalise according to a typical average mass-to-light relation established here, to obtain a reliable deflection field based on the distribution and luminosity of bright cluster members.

This procedure, as we have shown in many previous SL analyses, is sufficient to derive the critical curves with enough accuracy to immediately identify many multiple-images across the lensing field, as the primary mass distribution is initially well-constrained. Here we used a subsample of 10 well-studied clusters covered by both SDSS and HST to calibrate and test our analysis method, and showed that remarkably accurate determination of the Einstein radius can be made in an automated way, based on the light distribution of bright galaxies, and scaled by their luminosity. A tight correlation is seen between the Einstein radii derived in detailed analyses of HST data and using multiple images as input, and those from the ``blind'' survey tool presented here, operated on the same clusters but in SDSS data and without using any multiple-images information as input.

This efficient modelling method enables us to present the first observationally-based representative Einstein radius distribution, based on a coherent unbiased sample of the first 10,000 clusters in the Hao et al. (2010) catalog, larger by a few orders of magnitude than the number of SL clusters analysed to date. For this all-sky representative sample the Einstein radius distribution is log-normal in shape, with $\langle Log(\theta_{e}\arcsec)\rangle=0.73^{+0.02}_{-0.03}$, $\sigma=0.316^{+0.004}_{-0.002}$, and with higher abundance of large $\theta_{e}$ clusters than predicted by $\Lambda$CDM, and with a maximum of $\theta_{e}\simeq69\pm12 \arcsec$ ($z_{s}=2$) for the most massive candidate, in agreement with semi-analytic calculations.

In addition to characterising the overall Einstein distribution, we also uncover various relations with cluster properties listed in the probed catalog. For example, as may be expected, a clear relation is seen between the logarithmic Einstein radius distribution mean (for $z_{s}=2$), and the luminosity or richness, so that richer and more luminous clusters exhibit, on average, larger Einstein radii. An especially intriguing trend is found with the cluster redshift. On average, the mean effective Einstein radii steadily increase from $z\sim0.45$ to $z\sim0.1$. If real, and not fully accounted for by geometry (this requires knowledge of the mass profile, see \S \ref{dependency}), this may be a result of cluster evolution and relaxation processes, which make lower-$z$ clusters more concentrated, thus boosting the mass in the centre and thus the Einstein radius.

Reexamining the log-normal Einstein distribution in physical scales rather than angular scales, we obtain best-fitting values of $\langle Log(\theta_{e}~[$kpc$])\rangle=1.418\pm0.006$ and $\sigma=0.30\pm0.01$, for the full sample. Subdividing this distribution, the mean effective Einstein radii are constant throughout the volume-limited range of the catalog ($0.1<z_{l}<0.43$), $\langle Log(\theta_{e}~[$kpc$])\rangle=1.25\pm0.03$.

The redshift trend tentatively seen in our results could for instance be explained by a concentration-mass relation that evolves more steeply in redshift than found in numerical simulations. If confirmed, the redshift evolution indicated here could help deriving an observational concentration-mass relation once a broader calibration sample is available.

The results presented here are affected to some extent by statistical noise and uncertainty as detailed above (typically $\leq18\%$; \S \ref{results}), but it should be stressed that these uncertainties arise mostly from the data themselves, and not from the modelling method, which in light of higher-end data will produce much cleaner results. In fact, our SL algorithm could independently verify or at least probe, the cluster catalog and its cluster finding algorithm, by marking possible misidentified cluster candidates which do not follow the relations we obtained throughout this work.

Such an efficient modelling method can also aid in actually finding massive large lenses and many multiple-images, which in turn could be used to fine-tune the mass model and profile, especially when redshift information and preferably high-resolution deep space-imaging data are available. Combined with complementary data such as weak-lensing, this will allow for an extensive examination of many other cluster properties, such as the mass-concentration relation, or a ``universal'' mass profile shape (e.g., the CLASH program, Postman et al. 2011; see also Umetsu et al. 2011a,b). Further analysis results of the SDSS cluster catalog of Hao et al. (2010) will be presented in upcoming work, where we will aim to include a larger reference calibration sample to validate further the results and uncertainties presented here.

\section*{acknowledgments}

We thank the anonymous referee for significant and very constructive comments. AZ is grateful for the John Bahcall excellence prize which further encouraged this work, and to Sharon Sadeh, Dan Maoz, Irene Sendra, Gregor Seidel, Elisabeta Lusso, Bj\"orn M. Sch\"afer, Matthias Redlich, Joseph Hennawi and Andrea Macci\'o, for useful discussions.
This work was supported by contract research ``Internationale Spitzenforschung II/2''
of the Baden-W\"urttemberg Stiftung, and by the transregional collaborative research centre TR~33 of the Deutsche Forschungsgemeinschaft.
Part of this work is based on observations made with the NASA/ESA Hubble Space Telescope.
Funding for the SDSS and SDSS-II has been provided by the Alfred P. Sloan Foundation,
the Participating Institutions, the National Science Foundation, the U.S. Department of Energy,
the National Aeronautics and Space Administration, the Japanese Monbukagakusho, the Max
Planck Society, and the Higher Education Funding Council for England. The SDSS Web Site is
http://www.sdss.org/.
The SDSS is managed by the Astrophysical Research Consortium for the Participating Institutions.
The Participating Institutions are the American Museum of Natural History, Astrophysical
Institute Potsdam, University of Basel, University of Cambridge, Case Western Reserve University,
University of Chicago, Drexel University, Fermilab, the Institute for Advanced Study, the Japan
Participation Group, Johns Hopkins University, the Joint Institute for Nuclear Astrophysics, the
Kavli Institute for Particle Astrophysics and Cosmology, the Korean Scientist Group, the Chinese
Academy of Sciences (LAMOST), Los Alamos National Laboratory, the Max-Planck-Institute for
Astronomy (MPIA), the Max-Planck-Institute for Astrophysics (MPA), New Mexico State University,
Ohio State University, University of Pittsburgh, University of Portsmouth, Princeton University,
the United States Naval Observatory, and the University of Washington.
This work was supported in part by the FIRST program ``Subaru Measurements of Images and Redshifts (SuMIRe)'', World Premier
International Research Center Initiative (WPI Initiative), MEXT,
Japan, and Grant-in-Aid for Scientific Research from the JSPS
(23740161).

\bsp
\label{lastpage}


\begin{thebibliography}{}
\bibitem[]{}Abazajian, K., et al., 2003, AJ, 126, 2081
\bibitem[]{}Abazajian, K.N., et al., 2009, ApJS, 182, 543
\bibitem[]{}Bartelmann, M., Steinmetz, M., Weiss, A., 1995, A\&A, 297, 1
\bibitem[]{}Bartelmann, M., 2010, arXiv:1010.3829
\bibitem[]{}Bauer, A.H., Baltay, C.,; Ellman, N., Jerke, J., Rabinowitz, D., Scalzo, R., 2012, arXiv:1202.1371
\bibitem[]{}Bayliss, M.B., Gladders, M.D., Oguri, M., Hennawi, J.F., Sharon, K., Koester, B.P., Dahle, H., 2011a, ApJ, 727L, 26
\bibitem[]{}Bayliss, M.B., Hennawi, J.F., Gladders, M.D., Koester, B.P., Sharon, K., Dahle, H., Oguri, M., 2011b, ApJS, 193, 8
\bibitem[]{}Becker, M.R., et al., 2007, ApJ, 669, 905
\bibitem[]{}Ben\'itez, N., et al., 2009, ApJ, 691, 241
\bibitem[]{}Brada\v{c}, M., et al., 2005, A\&A, 437, 49
\bibitem[]{}Brada\v{c}, M., et al., 2006, ApJ, 652, 937
\bibitem[]{}Broadhurst, T., et al. 2005a, ApJ, 621, 53
\bibitem[]{}Broadhurst, T., Takada, M., Umetsu, K., Kong, X., Arimoto, N., Chiba, M., Futamase, T., 2005b, ApJ, 619, 143
\bibitem[]{}Broadhurst, T. \& Barkana, R., 2008, MNRAS, 390, 1647
\bibitem[]{}Broadhurst, T, Umetsu, K, Medezinski, E., Oguri,M., Rephaeli, Y., 2008, ApJ 685, L9
\bibitem[]{}Bullock J.S., Kolatt T.S., Sigad Y., Somerville R.S., Kravtsov A.V., Klypin A.A., Primack J.R., Dekel A., 2001, MNRAS, 321, 559
\bibitem[]{}Cabanac, R.A., et al., 2007, A\&A, 461, 813
\bibitem[]{}Coe, D., Ben\'itez, N., Broadhurst, T., Moustakas, L.A., 2010, ApJ, 723, 1678
\bibitem[]{}Coe, D., Fuselier, E., Ben\'itez, N., Broadhurst, T., Frye, B., Ford, H., 2008, ApJ, 681, 814
\bibitem[]{}Coe, D., et al., 2012, arXiv:1201.1616
\bibitem[]{}Corless, V.L. \&  King, L.J., 2009, MNRAS, 396, 315
\bibitem[]{}Dalal, N., Holder, G., Hennawi, J.F., 2004, ApJ, 609, 50
\bibitem[]{}Diego J. M., Sandvik H. B., Protopapas P., Tegmark M., Ben\'itez N., Broadhurst T., 2005, MNRAS, 362, 1247
\bibitem[]{}Donnarumma, A. et al., 2011, A\&A, 528A, 73
\bibitem[]{}Duffy, A.R., Schaye, J., Kay, Scott T.; Dalla Vecchia, C., 2008, MNRAS, 390L, 64
\bibitem[]{}Eisenstein, D.J., 2005, ApJ, 633, 560
\bibitem[]{}Flores, R.A., Maller, A.H., Primack, J.R., 2000, ApJ, 535, 555
\bibitem[]{}Gavazzi, R., Fort, B., Mellier, Y., Pell\'o, R., Dantel-Fort, M.,  2003, A\&A, 403, 11
\bibitem[]{}Giocoli, C., Meneghetti, M., Bartelmann, M., Moscardini, L., Boldrin, M., 2011, arXiv:1109.0285
\bibitem[]{}Gralla, M.B., et al., 2011, ApJ, 737, 74
\bibitem[]{}Grillo, C., Eichner, T., Seitz, S., Bender, R., Lombardi, M., Gobat, R., Bauer, A., 2010, ApJ, 710, 372
\bibitem[]{}Halkola A., Hildebrandt H., Schrabback T., Lombardi M., Brada\v{c} M., Erben T., Schneider P., Wuttke D., 2008, A\&A, 481, 65
\bibitem[]{}Halkola, A., Seitz, S., Pannella, M., 2006, MNRAS, 372, 1425
\bibitem[]{}Hao, J., et al., 2009, ApJ, 702, 745
\bibitem[]{}Hao, J., et al., 2010, ApJS, 191, 254
\bibitem[]{}Hennawi J.F., Dalal N., Bode P., Ostriker J.P., 2007a, ApJ, 654, 714
\bibitem[]{}Hennawi J.F., Dalal N., Bode P., 2007b, ApJ, 654, 93
\bibitem[]{}Hennawi J.F., et al., 2008, AJ, 135, 664
\bibitem[]{}Hilbert, S., White, S.D.M., Hartlap, J., Schneider, P., 2007, MNRAS, 382, 121
\bibitem[]{}Hildebrandt, H., et al., 2011, arXiv:1103.4407
\bibitem[]{}Horesh, A., Maoz, D., Hilbert, S., Bartelmann, M., arXiv:1101.4653
\bibitem[]{}Jing, Y. P. \& Suto, Y., 2002, ApJ, 574, 538
\bibitem[]{}Johnston, D.E., et al., 2007a, arXiv:0709.1159
\bibitem[]{}Johnston, D.E., Sheldon, E.S., Tasitsiomi, A., Frieman, J.A., Wechsler, R.H., McKay, T.A., 2007b, ApJ, 656, 27
\bibitem[]{}Jullo, E., Kneib, J.-P., Limousin, M., Elíasdóttir, A., Marshall, P. J., Verdugo, T., 2007, NJPh, 9, 447
\bibitem[]{}Keeton, C. R. 2001, preprint [astro-ph/0102340]
\bibitem[]{}Kneib, J.-P., Ellis, R. S., Smail, I., Couch, W. J., Sharples, R. M , 1996, ApJ, 471, 643
\bibitem[]{}Koester, B.P., et al., 2007, ApJ, 660, 239
\bibitem[]{}Komatsu, E., et al., 2011, ApJS, 192, 18
\bibitem[]{}Kubo, J.M., et al., 2010, ApJ, 724L, 137
\bibitem[]{}Liesenborgs, J., De Rijcke, S., Dejonghe, H., 2006, MNRAS, 367, 1209
\bibitem[]{}Liesenborgs, J., De Rijcke, S., Dejonghe, H., Bekaert, P., 2007, MNRAS, 380, 1729
\bibitem[]{}Liesenborgs, J., De Rijcke, S., Dejonghe, H., Bekaert, P., 2009, MNRAS, 397, 341
\bibitem[]{}Limousin, M., et al., 2007, ApJ, 668, 643
\bibitem[]{}Limousin, M., et al., 2008, A\&A, 489, 23
\bibitem[]{}Limousin, M., et al., 2010, MNRAS, 405, 777
\bibitem[]{}Limousin, M., et al., 2011, arXiv:1109.3301
\bibitem[]{}Macci\'o, Andrea V., Dutton, Aaron A., van den Bosch, Frank C., 2008, MNRAS, 391, 1940
\bibitem[]{}Mandelbaum, R., Seljak, U., Cool, R.J., Blanton, M., Hirata, C.M., Brinkmann, J., 2006, MNRAS, 372, 758
\bibitem[]{}Marshall, P.J., Hogg, D.W., Moustakas, L.A., Fassnacht, C.D., Brada\v{c}, M., Schrabback, T., Blandford, R.D., 2009, ApJ, 694, 924
\bibitem[]{}Maughan, B.J., Jones, C., Forman, W., Van Speybroeck, L., 2008, ApJS, 174, 117
\bibitem[]{}Meneghetti, M., Bolzonella, M., Bartelmann, M., Moscardini, L., Tormen, G., 2000, MNRAS, 314, 338
\bibitem[]{}Meneghetti, M., Fedeli, C., Pace, F., Gottl\"ober, S., Yepes, G., 2010a, A\&A, 519A, 90
\bibitem[]{}Meneghetti, M., Rasia, E., Merten, J., Bellagamba, F., Ettori, S., Mazzotta, P., Dolag, K., Marri, S., 2010b, A\&A, 514A, 93
\bibitem[]{}Meneghetti, M., Fedeli, C., Zitrin, A., Bartelmann, M., Broadhurst, T., Gottl\"oeber, S., Moscardini, L., Yepes, G., 2011, arXiv:1103.0044, A\&A in press
\bibitem[]{}Merten, J., Cacciato, M., Meneghetti, M., Mignone, C., Bartelmann, M., 2009, A\&A, 500, 681
\bibitem[]{}Merten, J., et al., 2011, MNRAS, 417, 333
\bibitem[]{}Moles, M., S\'anchez, S.F., Lamadrid, J.L., Cenarro, A.J., Crist\'obal-Hornillos, D., Maicas, N., Aceituno, J., 2010, PASP, 122, 363
\bibitem[]{}Narayan, R. \& Bartelmann, M., Lectures on Gravitational Lensing, 1996, arXiv:astro-ph/9606001v2
\bibitem[]{}Oguri, M. \& Blandford, R.D., 2009, MNRAS, 392, 930
\bibitem[]{}Oguri, M., et al., 2009, ApJ, 699, 1038
\bibitem[]{}Oguri, M. \& Takada, M. 2011, PhRvD, 83b, 023008
\bibitem[]{}Ponente, P.P. \& Diego, J.M., 2011, A\&A, 535A, 119
\bibitem[]{}Postman, et al., 2011, arXiv:1106.3328
\bibitem[]{}Prada, F., Klypin, A.A., Cuesta, A.J., Betancort-Rijo, J.E., Primack, J., 2011, arXiv:1104.5130
\bibitem[]{}Press, W.H. \& Schechter P., 1974, ApJ, 187, 425
\bibitem[]{}Puchwein, E. \& Hilbert, S., 2009, MNRAS, 398, 1298
\bibitem[]{}Richard, J., et al., 2010, MNRAS, 404, 325
\bibitem[]{}Richard, J., Pei, L., Limousin, M., Jullo, E., Kneib, J. P., 2009, A\&A, 498, 37
\bibitem[]{}Rozo, E., et al., 2009, ApJ, 699, 768
\bibitem[]{}Rozo, E., et al., 2010, ApJ, 708, 645
\bibitem[]{}Sadeh, S. \& Rephaeli, Y., 2008, MNRAS, 388, 1759
\bibitem[]{}Sand, D.J., Treu, T., Ellis, R.S., Smith, G.P., 2005, ApJ, 627, 32
\bibitem[]{}Seljak, U., et al., 2005, PhRvD, 71j3515
\bibitem[]{}Sereno, M.; Jetzer, Ph.; Lubini, M, 2010, MNRAS, 403, 2077
\bibitem[]{}Sheldon, E.S., et al., 2009, ApJ, 703, 2217
\bibitem[]{}Tegmark, M., et al., 2004, ApJ, 606, 702
\bibitem[]{}Tegmark, M., et al., 2006, PhRvD, 74l3507
\bibitem[]{}Torri, E., Meneghetti, M., Bartelmann, M., Moscardini, L., Rasia, E., Tormen, G., 2004, MNRAS, 349, 476
\bibitem[]{}Tremonti, C.A., et al., 2004, ApJ, 613, 898
\bibitem[]{}Umetsu, K., et al., 2009, ApJ, 694, 1643
\bibitem[]{}Umetsu, K., Broadhurst, T., Zitrin, A., Medezinski, E., Hsu, L.-Y., 2011a, ApJ, 729, 127
\bibitem[]{}Umetsu, K., Broadhurst, T., Zitrin, A., Medezinski, E., Coe, D., Postman, M., 2011b, arXiv:1105.0444
\bibitem[]{}Wambsganss, J., Cen, R., Ostriker, J.P., Turner, E.L., 1995, Sci, 268, 274
\bibitem[]{}Webster, R.L., Hewett, P.C., Irwin, M.J., 1988, AJ, 95, 19
\bibitem[]{}Wen, Z-L., Han, J-L., Jiang, Y.-Y., 2011, RAA, 11.1185
\bibitem[]{}Wojtak, R., Hansen, S.H., Hjorth, J., 2011, Nature, 477, 567
\bibitem[]{}Zitrin, A., Broadhurst, T., Rephaeli, Y., Sadeh, S., 2009a, ApJ, 707L, 102
\bibitem[]{}Zitrin, A., et al., 2009b, MNRAS, 396, 1985
\bibitem[]{}Zitrin, A., et al., 2010, MNRAS, 408, 1916
\bibitem[]{}Zitrin, A., Broadhurst, T., Barkana, R., Rephaeli, Y., Ben\'itez, N., 2011a, MNRAS, 410, 1939
\bibitem[]{}Zitrin, A., Broadhurst, T., Coe, D., Liesenborgs, J., Ben\'itez, N., Rephaeli, Y., Ford, H., Umetsu, K., 2011b, MNRAS, 413, 1753
\bibitem[]{}Zitrin, A., et al., 2011c, ApJ, 742, 117

%
\end{thebibliography}
\end{document}